\documentclass{emulateapj}

\slugcomment{Draft}

\shorttitle{ The VIRUS-P Exploration of Nearby Galaxies (VENGA)}
\shortauthors{Blanc et al.}

\begin{document}

\title{The VIRUS-P Exploration of Nearby Galaxies (VENGA): Survey
  Design, Data Processing, and Spectral Analysis Methods}

\author{Guillermo A. Blanc\altaffilmark{1}, 
Tim Weinzirl\altaffilmark{2}, 
Mimi Song\altaffilmark{2}, 
Amanda Heiderman\altaffilmark{2}, 
Karl Gebhardt\altaffilmark{2}, 
Shardha Jogee\altaffilmark{2}, 
Neal J. Evans II\altaffilmark{2}, 
Remco C. E. van den Bosch\altaffilmark{3}, 
Rongxin Luo\altaffilmark{7}, 
Niv Drory\altaffilmark{4}, 
Maximilian Fabricius\altaffilmark{5}, 
David Fisher\altaffilmark{6}, 
Lei Hao\altaffilmark{7}, 
Kyle Kaplan\altaffilmark{2},
Irina Marinova\altaffilmark{2}, 
Nalin Vutisalchavakul\altaffilmark{2}, 
Peter Yoachim\altaffilmark{8}}

\altaffiltext{1}{Observatories of the Carnegie Institution for Science, Pasadena, CA}
\altaffiltext{2}{Department of Astronomy, The University of Texas at Austin, Austin, TX}
\altaffiltext{3}{Max Planck Institute for Astronomy, Heidelberg, Germany}
\altaffiltext{4}{Instituto de Astronom\'ia, Universidad Nacional Aut\'onoma de M\'exico, M\'exico DF, M\'exico}
\altaffiltext{5}{Max Planck Institute for Extraterrestrial Physics, Garching, Germany}
\altaffiltext{6}{Department of Astronomy, University of Maryland, College Park, MD}
\altaffiltext{7}{Shanghai Astronomical Observatory, Shanghai, China}
\altaffiltext{8}{Astronomy Department, University of Washington, Seattle, WA}

\begin{abstract}

We present the survey design, data reduction, and spectral fitting pipeline
for the VIRUS-P Exploration of Nearby Galaxies
(VENGA). VENGA is an integral field spectroscopic survey, which maps
the disks of 30 nearby spiral galaxies. Targets span a wide range in
Hubble type, star formation activity, morphology, and
inclination. The VENGA data-cubes have $5.6''$ FWHM spatial resolution,
$\sim5$\AA\ FWHM spectral resolution, sample the 3600\AA-6800\AA\
range, and cover large areas typically sampling galaxies out to
$\sim0.7 R_{25}$. These data-cubes can be used to produce 2D maps of 
the star formation rate, dust extinction, electron
density, stellar population parameters, the kinematics and chemical abundances
of both stars and ionized gas, and other physical quantities derived
from the fitting of the stellar spectrum and the measurement of
nebular emission lines. To exemplify our methods and the quality of the data, we
present the VENGA data-cube on the face-on Sc galaxy \mbox{NGC 628}
(a.k.a. M 74).  The VENGA observations of NGC 628 are described, as
well as the construction of the data-cube, our spectral fitting method, and the
fitting of the stellar and ionized gas velocity fields. We also propose a
new method to measure the inclination of nearly face-on systems based
on the matching of the stellar and gas rotation curves using asymmetric
drift corrections. VENGA will measure relevant physical
parameters across different environments within these galaxies, allowing
a series of studies on star formation, structure
assembly, stellar populations, chemical evolution, galactic feedback,
nuclear activity, and the properties of the interstellar medium in massive
disk galaxies.

\end{abstract}


\section{Introduction}

In the context of $\Lambda$CDM cosmology, the formation and
evolution of galaxies takes place in the bottom of the deepest gravitational potential wells in the dark matter
distribution (dark matter halos). Gas accretion into these halos and merging processes
ultimately trigger star formation giving rise to galaxies
\citep{blumenthal84, white88, somerville99}. Although consensus has been reached concerning
this picture, the details of the baryonic physics behind galaxy formation in the
centers of dark matter halos are still aggressively debated. The triggering of star
formation and the variables that set the star formation rate
\citep[$SFR$, ][]{kennicutt98b, kennicutt12, mckee07, leroy08, krumholz09b}, the contribution
from different types of feedback processes \citep[AGN, supernovae, stellar
radiation, e.g. ][]{kauffmann99, croton06, thompson09}, as well as the impact
of gas accretion from the inter-galactic medium
\citep[IGM, e.g. ][]{dekel09, fumagalli11, dave11, vandevoort12}, at
regulating the gaseous budget, structure,
chemical composition, and kinematics of the ISM, and the role
that major and minor mergers as well as secular evolution processes
play at shaping galaxies \citep{toomre72, kormendy04}, are
currently the subjects of active research. All these processes play a major role in
determining how galaxies evolve throughout cosmic time, building up
their stellar mass and shaping their present day structure.
 
The detailed manner in which the above physical phenomena (star
formation, gas accretion, feedback, interactions, and secular
evolution) proceed, ultimately determines the morphology, kinematics,
and chemical structure of both the ISM and stellar components of a
galaxy. We can therefore study these processes by obtaining spatially
resolved measurements of quantities like the $SFR$, the metallicity, the stellar and
ionized gas velocity and velocity dispersion, the age of the observed stellar
populations, the atomic and molecular gas surface density, etc.,
and analyzing how they relate to each other and to the global
properties of the galaxies under study. In this way, we can test
current theoretical models describing the above phenomena and their
impact on galaxy evolution. 

Wide field optical integral field
unit (IFU) spectroscopy allows the measurement of many of these quantities in
nearby galaxies. Optical IFU data-cubes, in combination with
multi-wavelength broad band photometry, and \mbox{sub-mm} and radio maps of
the same galaxies, are powerful datasets to study galaxy evolution.

Integral field spectroscopy of nearby galaxies has been somewhat
limited in the past, mostly due to the small field-of-view
of available IFUs. During the last decade, a
new generation of wide field integral field spectrographs like SAURON
on the 4.2m William Herschel Telescope \citep{bacon01}, PPAK on the
3.5m at Calar Alto Observatory \citep{kelz06}, SparsePak on the WIYN
3.5m telescope \citep{bershady04}, and the Mitchell Spectrograph
(formerly known and hereafter referred to as VIRUS-P) on the 2.7m Harlan
J. Smith telescope at McDonald Observatory \citep{hill08a}, have
opened the path to study nearby systems subtending large angular
diameters on the sky. 

Early surveys of nearby galaxies using IFUs mostly focused on studying
the kinematics and stellar populations of early type systems. These
include the SAURON Survey \citep{bacon01, dezeeuw02}, and its
extension, the Atlas3D Survey \citep{cappellari11} which by now have
mapped hundreds of elliptical and lenticular galaxies. Wide field IFU
studies of later type disk galaxies include the work of
\cite{ganda06} who used SAURON to observe the central regions of 18
nearby late-type spirals, the Disk Mass Project \citep{bershady10}
which used SparsePak and PPAK to measure H$\alpha$ velocity fields for
146 face-on spirals, and stellar kinematics for a subset of 46
objects, with the aim of constraining the distribution of stellar mass
and dark matter in disk galaxies, and the PPAK IFS Nearby Galaxies
Survey \citep[PINGS,][]{rosales-ortega10}, which maps the disks of 17
nearby disk galaxies. The PPAK IFU is currently being used to conduct
the Calar Alto Legacy Integral Field Area survey \citep[CALIFA,
][]{sanchez11b}, a massive project mapping $\sim600$ galaxies of all
Hubble types, selected based on their angular size and distance (in
order for them to fill the PPAK field-of-view). In the future the
MANGA SDSS IV project\footnote{http://www.sdss3.org/future/} will
produce optical datacubes for tens of thousands of galaxies. A number of IFU
studies of galaxies have been conducted
at high redshift ($1<z<3$), where target sizes are well suited to the
small fields of view of IFUs in 10m class telescopes
\citep[e.g. ][]{genzel06, forster-schreiber06, law07, wright07,
  lemoine10}.

In this work, we present the VIRUS-P Exploration of Nearby Galaxies
(VENGA), an IFU survey of 30 nearby spirals which uses
VIRUS-P (currently the IFU with the largest field-of-view in the world) to
spectroscopically map large portions of the disks of these objects.
The sample spans a wide range in Hubble types, $SFR$s, and morphologies,
including galaxies with classical and pseudo-bulges as well as barred and
unbarred objects. Ancillary multi-wavelength data exist for many of
the targets. This includes HST optical and near-IR imaging with ACS
and NICMOS, Spitzer mid-IR and far-IR imaging with IRAC
and MIPS, near-UV and far-UV imaging from GALEX, and far-IR HERSCHEL
data. Also CO and HI 21cm maps are available for most of the
sample. VENGA's potential lies in a combination of wide spatial
and spectral coverage, good spatial resolution, and depth. The large $1.7'\times
1.7'$ field-of-view of the IFU allows us to typically sample each
system out to $\sim0.7R_{25}$ by tiling only a few VIRUS-P
pointings. The size of the VIRUS-P optical fibers
($4.24''$ in diameter) samples physical scales of $\sim$300 pc at the
median distance of our targets and makes our observations very
sensitive to low-surface brightness emission. We obtain spectra with
a median $S/N=40$ per fiber per spectral
resolution element which allows good measurements of
stellar absorption and nebular emission line spectral features at the
native spatial resolution of the instrument over most of the data-cube
for every galaxy.

In comparison to other IFU surveys of spiral galaxies VENGA is
factors of a few deeper than surveys like PINGS and CALIFA while
having a similar wavelength coverage and a factor fo two larger
spatial resolution. As mentioned above a key advantage of the VENGA
sample is the existance of extensive multi-wavelength ancillary data
for most of these very nearby galaxies. In particular the far-IR, mm, and radio data
is essential for studying the relation between stars and the ISM and
the process of star formation. While CALIFA and in the future MANGA
will produce datacubes for hundreds and tens of thousands of spiral
galaxies, only few of those targets will have such extensive ancillary
datasets. In that sense the VENGA data can provide a good base to
interpret results from larger but shallower surveys of more
distant galaxies.

In the near future, VENGA will also include a high spectral resolution
component consisting of IFU mapping at $R\sim6000$ of the central
regions of a subset of the sample using the VIRUS-W spectrograph
\citep{fabricius08}. The high resolution VIRUS-W observations will
complement the broad wavelength range VIRUS-P observations and will
provide a more detailed view of the kinematics of gas and stars in the
central regions of spiral galaxies. This component of the survey will
be presented in a future publication. In this work we limit ourselves
to discussing the VIRUS-P data.

The VENGA data are being used to conduct an extensive set of studies on
star-formation, structure assembly, stellar populations, gas and
stellar dynamics, chemical evolution, ISM structure, and galactic
feedback. The data will also provide an excellent local universe control
sample for IFU studies of high-z galaxies. The survey is designed with
the following science goals:

\begin{itemize}

\item Study the process of star-formation on galactic scales
  \citep[][and references therein]{kennicutt12},
  including the correlations between the $SFR$ and the star formation
  efficiency ($SFE$) with other parameters like gas and stellar surface
  density, metallicity, galaxy dynamics, and stellar
  populations. The ultimate goal is to understand what are the relevant
  parameters setting the $SFR$ across different environments within
  galaxies \citep[see][]{blanc09, blanc12}.

\item Investigate the assembly of central spheroidal stellar components
  in disk galaxies \citep[][and references therein]{kormendy04}. This includes characterizing the dynamics,
  stellar populations, and chemical abundances of classical
  and pseudo-bulges in spiral galaxies, in order to constrain their star-formation history and
  understand their origin. The goal is to distinguish between different
  evolutionary paths that might give rise to these structures (secular
  evolution, galaxy-galaxy interactions, etc.).

\item Provide detailed observations of radial gas inflow in
  the central parts of disk galaxies, induced by barred and oval
  potentials \citep[][and references therein]{kormendy04}. This includes studying the velocity field of ionized gas
  and stars in the regions influenced by the presence of bars and
  ovals, and also the effects of induced gas condensation and shocks
  in the local star formation efficiency.
 
\item Construct two-dimensional maps of the stellar and gas phase
  metallicity of spiral galaxies \citep[e.g.][]{moustakas10, rosales-ortega11}. These maps will allow the study of
  radial abundance gradients measured with exquisite detail,
  the dispersion in abundances as a function of galactocentric radius, and
  deviations from axisymmetry in the two dimensional metallicity
  distribution. Comparing these measurements to chemical evolution
  models will help constrain the chemical enrichment, gas accretion, and
  star-formation history of disk galaxies in the local universe.

\item Use nebular emission line diagnostics to study the nature of
  the ionizing sources influencing different parts of the disks of
  spirals \citep[e.g.][and references therein]{kewley01, ho08,
    haffner09}. This includes the study of low luminosity active
  galactic nuclei (LLAGN), and their impact on the physical conditions of gas
  in the central parts of galaxies, and the
  study of the structure, metallicity, and kinematics of the diffuse
  ionized gas (DIG) in the ISM. In particular, for the two edge-on
  systems in the sample, we will also be able to
  constrain the properties of the DIG as a
  function of distance above the mid-plane. All these
  studies will provide insight about different feedback
  processes at play in star-forming disk galaxies, and the transport of
  UV photons in the multi-phase ISM.

\item Study the distribution of stellar mass and dark matter in
  spiral galaxies, by using a combination of the VENGA gas and stellar
  velocity fields and constraints on the $M/L$ ratio from stellar
  populations \citep[e.g.][]{bershady10, adams12}. The data should allow us, in principle, to set
  constraints on the shape of the dark matter halo density profile on
  these systems, which we can use to test the predictions of
  $\Lambda$CDM models. Of particular importance for these measurements
  is the high spectral resolution VIRUS-W component of the survey.

\end{itemize}

Throughout this paper we use the VENGA data on the face-on SA(s)c galaxy
NGC 628 to exemplify our methodology. This nearby grand-design spiral
galaxy has been extensively studied in the past and IFU spectroscopy
has been obtained in its central region by \cite{ganda06} and over a
larger area than presented here, although to a shallower depth, by
the PINGS survey \citep{rosales-ortega11, sanchez11b}. In this work we limit ourselves
to presenting the methods used to observe, reduce, and analyze the
VENGA IFU data, and we postpone the presentation of scientific results
on NGC 628 to separate publications, including an accompanying paper
studying the radial profile of the CO to H$_2$ conversion factor
($X_{CO}$) and its relation to the physical conditions across the disk
of the galaxy \citep{blanc12}. We also expect to present and make
public the full VENGA dataset in a future publication once the processing and
analysis of the whole survey is complete.

In \S2 we describe the survey design, including a description of the
VENGA sample and the observing strategy adopted to conduct the
survey. We present the VIRUS-P observations of NGC 628 in \S3. The
data processing pipeline and construction of the
final VENGA data products (i.e. reduced and calibrated spectral
data-cubes) is presented in \S4, followed by a description of the
techniques used to fit the spectra, measure stellar and gas
kinematics, and extract emission line fluxes (\S5). In \S6 we
fit the stellar and ionized gas velocity fields of NGC 628 using a harmonic
decomposition technique and introduce a new method to measure  the
inclination of nearly face-on systems based
on the matching of the stellar and gas rotation curves using asymmetric
drift corrections. Finally, we present our conclusions in
\S7. Throughout the paper we adopt a standard set of $\Lambda$CDM
cosmological parameters, $H_o=70$ km s$^{-1}$Mpc$^{-1}$,
$\Omega_M=0.3$, and $\Omega_{\Lambda}=0.7$ \citep{dunkley09}.

\begin{figure*}
\begin{center}
\epsscale{1.0}
\plotone{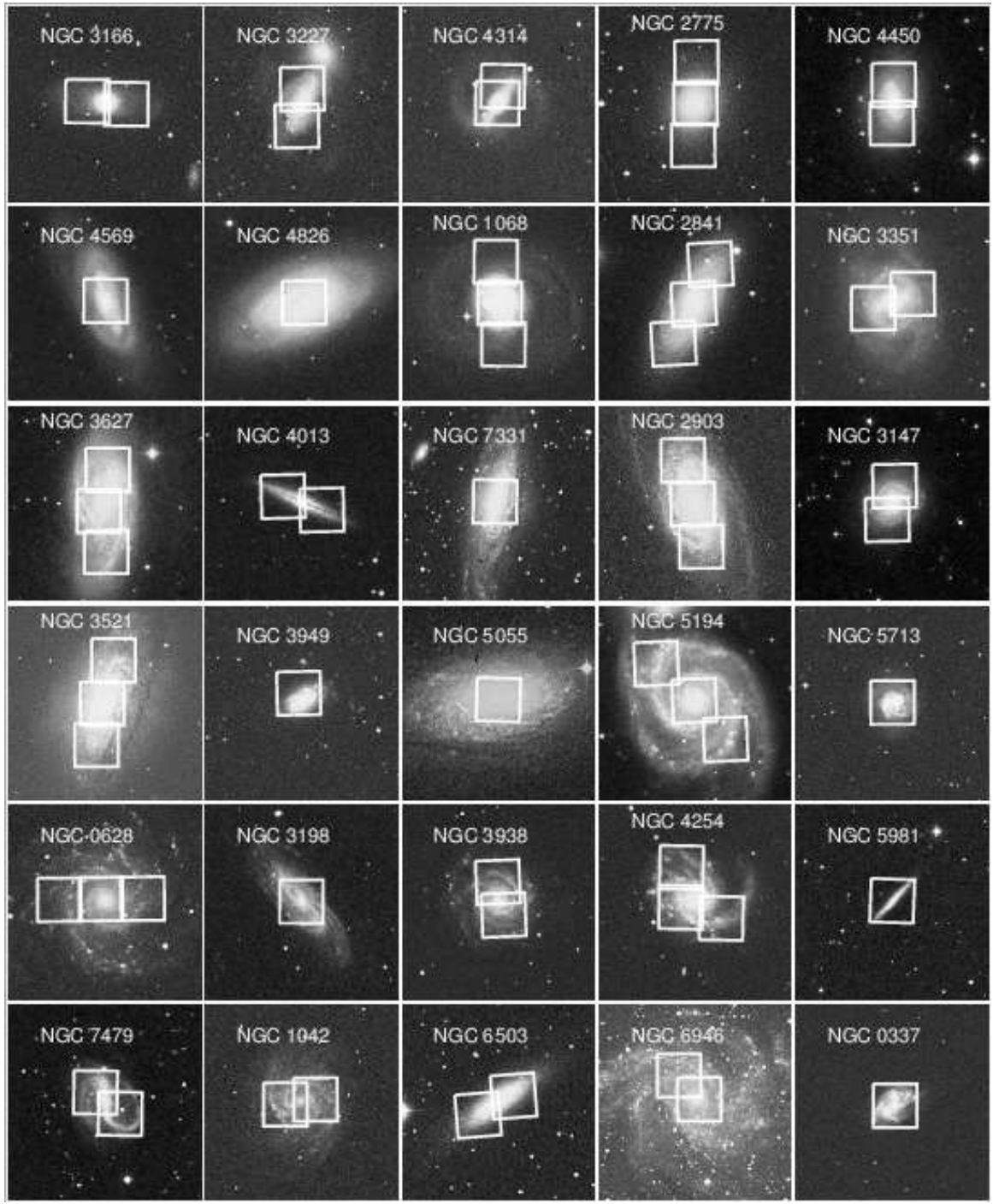}
\caption{Digital Sky Survey cutouts of the 30 galaxies in the VENGA
  sample. The targets are ordered by Hubble type (taken from RC3, \cite{devaucouleurs91}) from earlier to
  later. White boxes show the VIRUS-P $1.7'\times1.7'$ pointings
  observed on each galaxy.}
\label{fig-1}
\end{center}
\end{figure*}

\section{Survey Design}

In this section we present and characterize the VENGA sample and we
discuss the physical properties of the target galaxies. We also describe the observing strategy
and the instrumental configurations used to execute the survey.

\subsection{The VENGA Sample}

Table \ref{tbl-1} presents the galaxies observed in VENGA and lists
their main properties. Targets are
chosen to span a wide range in Hubble types, from S0 to Sd, a wide
range in inclinations from face-on to edge-on systems, and they include
both barred and unbarred objects. The sample also spans a wide range
in right ascension in order to allow observations to be carried out
throughout the whole year, and all objects have declinations
$\delta>-10^\circ$ to make them accessible from McDonald
Observatory. Figure \ref{fig-1} presents Digitized Sky Survey
(DSS\footnote{The Digitized Sky Surveys were produced at the Space
  Telescope Science Institute under U.S. Government grant NAG
  W-2166. The images of these surveys are based on photographic data
  obtained using the Oschin Schmidt Telescope on Palomar Mountain and
  the UK Schmidt Telescope. The plates were processed into the present
  compressed digital form with the permission of these
  institutions. }) cutouts for all the galaxies in VENGA. Overlaid are
the $1.7'\times1.7'$ VIRUS-P pointings obtained on each galaxy. While
VENGA is designed to map galaxies out to $\sim0.7R_{25}$, for
NGC 3198, NGC 4569, NGC 4826, NGC 5055, and NGC 7731, only a central
pointing was observed due to observing time constraints.

Since one of the goals of VENGA is to study the origin and properties
of stellar spheroids in the inner parts of disk galaxies, we included
objects showing both classical bulges and pseudo-bulges
\citep[][and references therein]{kormendy04}. To distinguish between
these two types of stellar structures we adopt a criterion based on
the S\'ersic index of the spheroidal component ($n_B$). Following the results of
\cite{fisher08} we adopt a limit of $n_{B}=2$. We consider classical
bulges those with $n_{B}>2$ and pseudo-bulges those with with
$n_{B}<2$. Table \ref{tbl-2} presents the bulge-to-total light
fractions (B/T) and $n_{B}$ values for 18 of the 30 VENGA galaxies
taken from \cite{dong06}, \cite{fisher08}, and \cite{weinzirl09}. Of the
galaxies for which we found bulge S\'ersic index measurements in the
literature, 50\% are classified as classical bulges and 50\% as
pseudo-bulges using the $n_{B}=2$ criterion.

In order to understand how well the galaxies in our sample represent the
overall population of galaxies in the local universe, we
compare their stellar mass ($M_*$) and $SFR$ distributions to that of $\sim8\times10^5$
galaxies at $z<0.2$ from the Sloan Digital Sky Survey
\citep[SDSS, ][]{york00}
MPA/JHU DR7\footnote{http://www.mpa-garching.mpg.de/SDSS/}
catalog of star formation rates \citep{brinchmann04}. For the VENGA
galaxies, we estimate $M_*$ from their $K$-band luminosity
by assuming a mass-to-light ratio of $\Upsilon_{K}=0.78$
\citep{dejong96}. The \mbox{$K$-band} luminosities are computed using the
distances reported in Table \ref{tbl-1} and the total $K$-band
apparent magnitudes from the 2MASS Large Galaxy Atlas
\citep[LGA, ][]{jarrett03} except for NGC 1042, NGC 3147, NGC 3949,
NGC 5981,
NGC 7479, and NGC 7331 which are not included in the LGA so their
magnitudes are taken from the 2MASS Extended Source Catalog
\citep{jarrett00}. We do not correct the luminosity for dust
extinction but we expect this effect to be small ($\sim$10\%
in the $K$-band). 

\begin{figure}[t]
\begin{center}
\epsscale{1.2}
\plotone{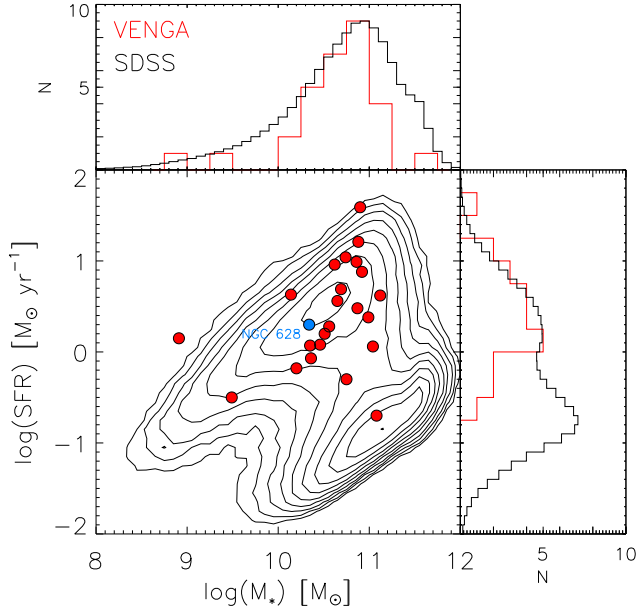}
\caption{Stellar mass versus star formation rate for the VENGA
  galaxies with $SFR$ measurements in Table \ref{tbl-3} (red circles),
and star forming galaxies in the SDSS MPA/JHU DR7 catalog (black
contours). The red and black histograms show the distributions for the
VENGA and SDSS galaxies respectively. The stellar mass histogram
includes the VENGA targets without $SFR$ measurements. NGC 628 is
shown as a blue circle.}
\label{fig-2}
\end{center}
\end{figure}

We were able to find integrated $SFR$ measurements in
the literature for 24 of the 30 galaxies in the VENGA sample. These
are taken from \cite{lee09}, \cite{kennicutt03}, and \cite{thilker07} in
that order of preference when multiple measurements are found. The
stellar masses and $SFR$s of the VENGA galaxies are reported in Table
\ref{tbl-3}. Figure \ref{fig-2} shows the VENGA and SDSS galaxies
on the $M_*$ versus $SFR$ plane. Our sample spans a range in $SFR$ from 0.2 to 39
M$_{\odot}$ yr$^{-1}$, and is distributed in this parameter similarly to
the star forming SDSS galaxies. In term of stellar mass, the VENGA galaxies
span a range between $8\times10^8$ M$_{\odot}$ and $3\times10^{11} $M$_{\odot}$, but with
93\% (28/30) of the sample having $M_*>10^{10}$ M$_{\odot}$. By comparing the
stellar mass distribution to that of the SDSS star forming galaxies it
is evident that our sample is biased towards the high-mass end of the
local star forming population. Figure \ref{fig-2} shows that our targets fall both
on and below the $M_*-SFR$ main sequence of star forming galaxies, so
starburst systems are not present in the sample. The VIRUS-P Investigation of the
Extreme Environments of Starbursts (VIXENS) project is conducting a
similar study to VENGA but on a sample of 15 interacting/starburst
galaxies in the local universe \citep{heiderman12}. VENGA is primarily a survey
of massive spiral galaxies. Dwarf galaxies, early type galaxies, and
starburst are not well represented in the sample.

\begin{figure}[t]
\begin{center}
\epsscale{1.2}
\plotone{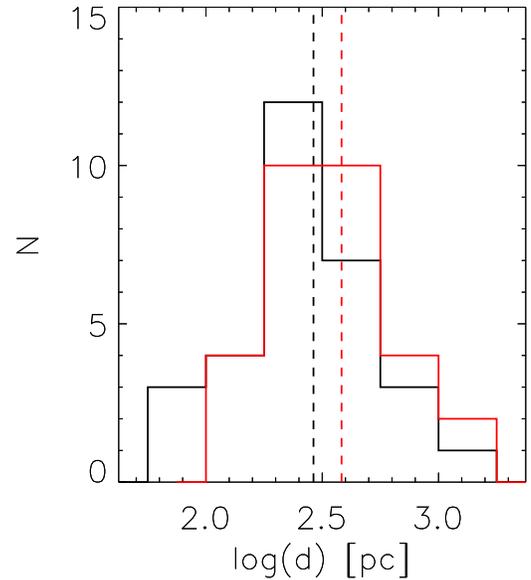}
\caption{Histogram of the logarithm of the VIRUS-P $4.24''$ fiber size in
  physical units for each galaxy in the VENGA sample (black), given
  the distances adopted in Table \ref{tbl-1}. Also shown is the size
  of the $5.6''$ FWHM PSF of the final data-cubes (red).The vertical dashed
lines marks the median native and effective spatial resolutions of 290
pc and 383 pc.}
\label{fig-3}
\end{center}
\end{figure}

 Finally, it is important to characterize the spatial resolution we can
achieve with VIRUS-P at the distances of our targets. Table
\ref{tbl-1} reports the physical scale in pc corresponding to one
arcsecond at the distance of each galaxy. The IFU fibers
subtend a $4.24''$ diameter on sky which corresponds to a spatial
resolution between 80 pc for the closest galaxy (NGC 6503) and 1020 pc for the
farthest (NGC 5981). As will be described in \S4.7 the PSF of the
final data-cubes is given by the convolution of the native
instrumental resolution (set by the fiber size) with the seeing and a
gaussian weighting filter used to combine different
frames at different dither positions. This translates into
an effective PSF FWHM of $5.6''$ in the final VENGA data-cubes which
corresponds to a range in spatial resolution of \mbox{109 pc - 1.35 kpc,}
and a median resolution of 383 pc. Figure \ref{fig-3} presents a
histogram of the physical scale in pc of the VIRUS-P native fiber size
(black) and the final data-cube PSF FWHM (red) for our targets.

\subsection{Observing Strategy}

The VIRUS-P IFU is a square array of 246 optical fibers (each $4.24''$
in diameter) which samples a $1.7'\times 1.7'$ field-of-view (FOV) with
a 1/3 filling factor. Three dithers provide full coverage of the FOV. 
The instrument images the spectra of the 246 fibers on a 2k$\times$2k
Fairchild Instruments charged-coupled device (CCD) with 15 $\mu$m
pixels. The CCD electronics deliver 3.6-4.3 e$^-$ read-noise depending
on the read-out mode used. Fiber spectra have an approximately gaussian spatial
profile of $\sim$5 pixels FWHM, and are stacked vertically on
the CCD at approximately 8 pixels apart from each other, minimizing
cross-talk between fibers.

For each \mbox{VIRUS-P} pointing we observe 3 dithers at relative positions
$(\Delta \alpha, \Delta \delta)=(0.0'',0.0'')$, $(-3.6'',-2.0'')$, and $(0.0'',
-4.0'')$ from the origin to ensure full coverage of the
FOV. Therefore in each pointing we obtain
spectra for 738 independent regions. 
Depending on the angular size of the targets we observe up to three
pointings on each galaxy
providing full coverage of the central part of the galaxies and a sampling
of the disk of most targets out to $\sim0.7 R_{25}$ (see Figure
\ref{fig-1} and last column of Table \ref{tbl-1}). Overall the VENGA
survey consists of 60 individual pointings composed of 3 dithers
each, amounting for spectra of $\sim44,000$ independent regions
(typically a few 100 pc in diameter) across the disks of the 30
galaxies in the sample.

The spectral range of VIRUS-P can be adjusted within the
3600-6800 \AA\ window. The instrument has a set of volume phase holographic
gratings which provide different spectral resolutions and wavelength
coverages. For VENGA we use the lowest resolution grating (VP1) which
provides a spectral resolution of $\sim5$ \AA\ FWHM and coverage over a
$\sim$2200 \AA\ wide spectral window
. We observe each galaxy in a blue setup
(3600-5800 \AA) and a red setup (4600-6800 \AA), obtaining
full spectral coverage in the 3600-6800 \AA\ range and doubling
the effective exposure time in the overlap region (4600-5800 \AA). All the data are
taken with $1\times1$ binning, which translates into a spectral
dispersion of $\sim1.11$ \AA\ pixel$^{-1}$, except for some early
observations of the central pointing of NGC 5194 which were taken using
$2\times1$ binning in the spectral direction \citep{blanc09}.

In terms of depth, the goal of VENGA is to obtain spectra that reaches
a median $S/N\sim40$ in continuum per spectral resolution element (FWHM) per
fiber across each galaxy. For most scientific applications this high
S/N per fiber requirement prevents us from having to bin our data over
many spatial resolution elements therefore ensuring good spatial resolution throughout
most of the maps. Exposure times for different galaxies were scaled using their effective $B$-band surface brightness
taken from the RC3 catalog \citep[][Table  \ref{tbl-1}]{devaucouleurs91} and typically range from 45 min to
3 hr per dither. The three dither positions and two wavelength setups
used imply effective exposure times ranging from 4.5 hr to 18 hr per
pointing.

\section{Observations}

The VENGA survey is still in the phase of data
acquisition. Observations of all targets in the red spectral setup
started in April 2008 and were completed in July 2010. Blue setup
observations started in September 2010 and we expect them to be 
completed during 2013. Table \ref{tbl-4} lists all the
observing runs we have conducted as part of VENGA, the instrumental
setup used, the number of nights observed, and the galaxies for which
data were obtained. 

As mentioned above the typical exposure times for each
dither range from 45 min to 3 hr, typically divided in shorter
15 to 25 min exposures. When conditions are not photometric we increase the exposure
times to ensure reaching the desired depth. Because
of the large angular diameter of our targets, during most observations
the \mbox{VIRUS-P} IFU never samples regions of blank sky. Therefore, off
source sky exposures are necessary to measure and subtract the sky
spectrum from the science data. We obtain 5 min sky frames bracketing
each science exposure. The off-source frames are taken 30' north of
each galaxy in fields that have been confirmed not to contain extended sources.

Bias frames, arc lamps, and twilight flats are obtained at the
beginning and end of each night. For the red setup we use a
combination of Ne+Cd comparison lamps and for the blue setup we
use Hg+Cd. These combinations of lamps provide a good set of strong lines over
the full spectral range of each setup allowing a good wavelength
calibration with minimal extrapolation towards the CCD edges.

During most nights we obtain data for one or two
spectro-photometric standard stars using the six-position fine
dithering pattern presented in \cite{blanc09}. As described in \S4.5
and \S4.6 standard stars are used to perform the relative flux
calibration while the absolute flux level is calibrated against
broad-band images. During some observing runs the spectra of 1 to 3 out of the 246
fibers fell off the CCD due to camera and grating alignment problems.
This translates in a lack of spectra for a few fibers in the corner of
the field-of-view which does not affect the data significantly.

\subsection{NGC 628 Data}

In this paper we present the VENGA data over 3
\mbox{VIRUS-P} pointings on the face-on Sc galaxy NGC 628. A summary
of the data are provided in Table \ref{tbl-5}. Observing conditions
were variable between different runs and within different nights during the same
observing run, ranging from photometric to partly cloudy conditions
with average atmospheric transparency down to $\sim$60\%. The seeing (as
measured from the guide star in the co-focal guider
camera of the instrument) ranged between $1.3''$ and $4.0''$
(FWHM) with a median of $2.0''$. 

Table \ref{tbl-5} presents a summary of the NGC 628 VENGA data after
rejecting 12 out of 207 frames (6\%) which are affected by pointing
errors or catastrophic sky subtraction problems (see \S4.4
and \S4.6). For each dither in each pointing on the galaxy we list
the total on-source exposure time, the number of frames, the average
seeing of the frames, and the mean atmospheric transparency (as
measured in \S4.6). Overall we obtained 84.2 hours of exposure on NGC
628 which translates into an extremely deep spectral data-cube for
this galaxy. The VENGA data on NGC 628 is roughly a factor of two
deeper than for the typical target in VENGA.

\section{Data Reduction and Calibration}

Data reduction is performed using the VACCINE pipeline
\citep{adams11a} in combination with a series of custom
built IDL routines. In this section we describe the data
processing including the flat-fielding, wavelength calibration, and
sky subtraction of individual science frames, and their 
flux calibration (both in the relative, i.e. across wavelength, and
absolute sense). We also discuss astrometric corrections applied to
the data. Both the astrometry and absolute flux calibration are based
on the comparison of reconstructed broad-band images from the VIRUS-P IFU spectra
and archival broad-band images of the galaxies. Finally we describe
the method used to combine individual frames taken in different
wavelength setups and at different positions in order to create the
final data-cubes and row-stacked-spectra (RSS) files for the VENGA
galaxies. At each step in the data reduction 
VACCINE and our set of external IDL routines propagate a properly
calculated error frame which is associated with each science frame
and used to compute the uncertainty in the final
spectra.

Throughout this section we use the NGC 628 VENGA data to exemplify the
results of different procedures. We describe processing steps in the
order in which they are applied to the data.

\subsection{Basic CCD Processing, Cosmic Ray Rejection, and Fiber Tracing}

All individual frames (bias, flats, arcs, sky, and science) are
overscan subtracted. We combine all the overscan subtracted bias
files for each observing run (usually $\sim100$) to create an image of
any residual bias structure which is subtracted from all the flat,
arc, sky, and science frames. 

We use the LA-Cosmic laplacian cosmic ray identification algorithm
of \cite{vandokkum01} to identify and mask cosmic rays in the science
images. We tune the algorithm to be robust enough to identify most
cosmic rays in the science frames while ensuring that real emission lines
are not masked. Any residual cosmic rays not identified in this
pass are removed from the data during the combination
of multiple frames (\S4.6).

Twilight flats are used to trace the spectrum of each fiber on the
CCD. VIRUS-P is mounted on a gimbal attached to the broken Cassegrain
focus of the 2.7m telescope. The gimbal keeps the spectrograph at a
constant gravity vector, making flexure effects on the optical path of
the instrument negligible. Thanks to this, with VIRUS-P there is no
need to obtain calibrations at the same time and telescope position of
the science data. We have observed shifts in the positions of
fibers on the CCD (at the 0.1 pixel level) when large changes in
temperature occur. VACCINE corrects for these small offsets during
the flat-fielding stage to properly remove the spatial PSF of
each fiber from the 2D spectra (\S4.3). To minimize
the magnitude of these temperature induce offsets we use the set of twilight-flats which is closest in
temperature to the average temperature at which the science data are
taken for both tracing and flat-fielding. Most of the time this means
the twilight-flats taken at dawn are used.

We extract the 2D spectrum of each fiber in the
science, sky, arc, and flat frames using a 5 pixel aperture centered
around the pixel containing the centroid of the fiber's spatial
profile. Using a discrete pixel instead of a fractional pixel aperture
centered on the trace centroid itself avoids having to re-sample
the data and conserves the noise properties of individual
pixels. Using a fixed aperture that matches the FWHM of the fiber
spatial profile implies that we recover 76\% of the total flux passing
through the fiber and makes the effects of neighboring fiber
cross-talk negligible. We prefer to adopt this approach instead of a
more sophisticated method to recover the flux in the wings of the
fiber spatial profile \citep[e.g. Gaussian Suppression,][]{sanchez06}
as the potential $\sim15$\% improvement in S/N that could be
gained in the case of perfect deblending and extraction is
hampered by the systematic flat-fielding uncertainty introduced by the
correction of temperature induced offsets in the edges of the fiber profile.

At this stage in the reduction VACCINE constructs a formal error map which
accounts for both read-noise and Poisson uncertainty for each pixel in
the 2D spectrum of each fiber. These maps are properly propagated
throughout the rest of the reduction (assuming gaussian uncertainties)
and end up providing the weights used when combining and collapsing
spectra from different frames, as well as the flux error in the final data-cube.

\subsection{Wavelength Calibration and Characterization of the
  Instrumental Spectral Resolution}

Arc lamp frames are combined to produce a master
arc for each night. VACCINE is typically able to match and fit $\sim12$
emission lines in the blue setup lamps (Hg+Cd) and $\sim20$ lines
in the red setup lamps (Ne+Cd). We fit the wavelength solution for
each fiber independently using a 4th order polynomial.
Residuals in the wavelength solution show an r.m.s. dispersion
of $\sigma_{\lambda}\simeq0.1$ \AA, or a tenth of a pixel ($\sim6$ km
s$^{-1}$ at 5000\AA).

We also use the emission lines in the master arc frame to characterize
the spectral resolution as a function of wavelength for each fiber. We
measure the FWHM of non-blended arc lines by performing single gaussian
fits. By fitting a second order polynomial to the measured FWHM values
across the spectral direction for each fiber we create a robust map of the
instrumental spectral resolution (FWHM$_{ins}$) of each fiber as a
function of wavelength. Good knowledge of the resolution is essential
at the time of fitting galaxy spectra with linear combinations of
empirical or synthetic templates which must be convolved to the same
resolution of the data in order to extract meaningful line-of-sight
velocity distributions (LOSVDs) from the fits.

The VIRUS-P instrumental resolution is observed to change
smoothly as a function of position on the detector, with values ranging
from 4.6\AA\ to 6.1\AA\ FWHM.  No variation in
the instrumental spectral resolution is observed between observing runs.
In the final data-cube of NGC 628 the median instrumental resolution in
velocity units is $\sigma_{inst}=123$ km s$^{-1}$.

\subsection{Flat Fielding}

The flat-fielding process in VACCINE is used to divide out three different
effects from the data: (1) the relative fiber-to-fiber throughput, (2)
the profile of the fiber PSF on the detector across the spatial
direction, and (3) the CCD pixel-to-pixel
variations in quantum efficiency. A master twilight flat is created
for each night by combining the set of flat frames closer in
temperature to the average throughout the night. First, VACCINE removes the signal
from the solar spectrum from the master twilight flat. For
each fiber, this is achieved by fitting a bspline to the
twilight spectra of a set of 60 neighboring fibers on the chip (which share a
similar spectral resolution) to create a template solar spectrum
(which also includes the spectral response of the instrument), and
then normalizing the spectrum of the fiber of interest
by this combined high S/N template \citep[see ][]{adams11a}. Since
each fiber provides an independent wavelength sampling of the twilight
spectrum, combining data from a large set of neighboring fibers
effectively yields a sub-pixel sampled spectrum which after
being fit by the bspline can be evaluated at the exact wavelength
scale of the fiber of interest. 

From the resulting normalized flat, VACCINE creates a ``fiber profile'' flat by
running a median smoothing kernel across the spectral direction. This
new frame contains only the relative fiber-to-fiber throughput and the
fiber PSF spatial profile. Dividing the normalized flat by this smoothed version
yields a pixel-to-pixel flat that is applied to the data. 

As mentioned in \S4.1, small (sub-pixel) temperature induced offsets
in the fiber positions on the CCD are sometimes present in the
data. This can translate into undesired residuals when removing the spatial profile
of the fibers if the data are divided by a flat-field that is offset
in the spatial direction. Residuals can be particularly large at
the edges of fibers, where the data values are divided by smaller numbers than at the
fiber core. These offsets must be corrected for, in order to remove
the fiber PSF across the spatial direction. To do so, VACCINE traces each fiber
in the science frame (after running a 30 pixel boxcar filter
across the spectral direction to ensure a high S/N measurement of the
fiber centroid), and computes an offset with respect to the same fiber
centroid in the ``fiber profile'' flat. These offsets are used to resample the smoothed ``fiber
profile'' flat using an optimal sinc-interpolation method in order to align
it with the data. We divide the science and sky frames by this
resampled smoothed flat frame, therefore removing the fiber spatial profile
from the data. 

\subsection{Sky Subtraction}

Sky spectra are measured by combining information from the two
off-source sky exposures taken before and after each science frame. In
\cite{blanc09} we simply averaged the before and after off-source frames to
create the sky frame used for background subtraction. While this
method worked well on the NGC 5194 data presented there, those
observations were taken far from twilight and under very stable
and dark conditions. We find that when conditions are not
optimal (e.g. close to twilight, when clouds are present, or near
moon-rise or moon-set) the sky brightness can change non-linearly with
time, making the simple averaging of bracketing sky frames
insufficient to produce an adequate sky subtraction. We adopt a
more sophisticated method to estimate and subtract the sky spectrum
from our data, which makes use of the temporal information we can
extract regarding the variability of the sky brightness as a function
of wavelength, from all the sky frames obtained throughout a single night.

\begin{figure}
\begin{center}
\epsscale{1.0}
\plotone{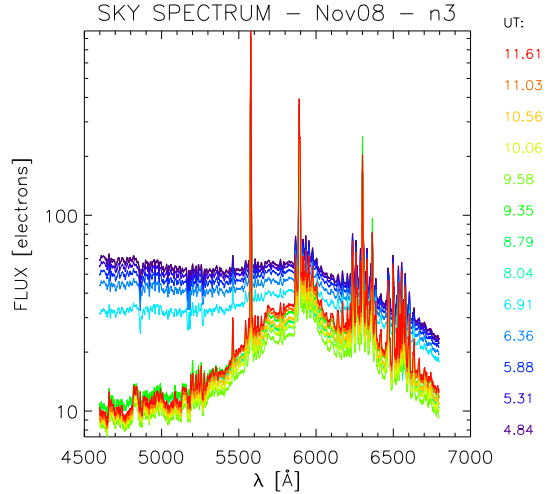}
\caption{Sky spectrum in raw units (before flux calibration) at
  different UT times (color coded) during the night of November 7th
  2008 for the VIRUS-P red setup.}
\label{fig-4}
\end{center}
\end{figure}

As an example, Figure \ref{fig-4} shows the raw (i.e. not flux
calibrated) sky spectrum measured from 13 off source sky frames
taken during the night of November 7th 2008 using the red
VIRUS-P setup. The spectra are color coded by UT time, with purple at
the beginning of the night and red at
the end of the night. We adopt the UT time corresponding to the middle
of each exposure. The elevated brightness and blue color of the
spectrum at the beginning of the night is due to the first quarter
moon. Moonset at McDonald Observatory on that date occurred at 8.47
hr UT and can be clearly seen as a sharp drop in the sky brightness,
particularly at blue wavelengths. For this night, the darkest skies
occurred between 9 hr and 10 hr UT. These are followed by a monotonous
increase in sky brightness which is steeper at redder
wavelengths. This brightening and reddening is due to both the
approachment of twilight and the fact that the observations were being
done at increasing airmass. Our sky subtraction method uses all this
information regarding the wavelength and time dependance of the sky
brightness throughout each night to correct the two bracketing
off-source sky exposures so they match the sky spectrum at the time
the science frame is taken.

\begin{figure}[t]
\begin{center}
\epsscale{1.0}
\plotone{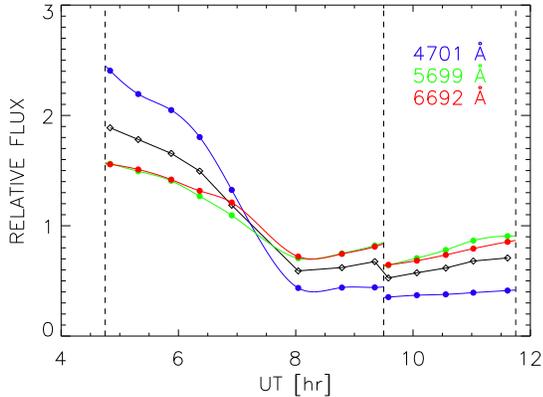}
\caption{Relative sky brightness as a function of UT time for the same
night shown in Figure \ref{fig-4} at three different wavelengths
(blue, green, and red). Filled circles correspond to measurements of
the sky brightness from the off-source background frames. The
beginning and end of observations of the same target are shown as
vertical dashed lines. Solid color curves show cubic spline fits to
the sky brightness. The black open diamonds and black solid curve show
the relative sky brightness averaged over the full spectrum.}
\label{fig-5}
\end{center}
\end{figure}

To trace these changes we average the spectra of
all fibers in each individual sky frame and then divide this average sky spectrum into 500
wavelength bins (each corresponding roughly to a
spectral resolution element). On a night-by-night basis, we use the
time stamps of all sky frames to construct a nightly ``light curve'' of the
night sky surface brightness at the central wavelength of each of
these 500 bins. Figure \ref{fig-5} shows the relative change in sky brightness
across the same night presented in Figure 4, for three different wavelength bins near
the blue end, middle, and red end of the spectrum (blue, green, and
red respectively). Measurements for
individual sky frames are shown as filled circles. For reference, the black
open diamonds show the change in bolometric brightness (i.e. in the
total flux integrated across the whole spectrum). The trends described in the last paragraph are clearly
seen. At the beginning of the night the sky brightness falls more steeply in the blue than
in the red as the moon sets, and rises faster in the red than in the
blue as we approach twilight at the end of the night. Vertical dashed lines mark the beginning and end of
observations of the same galaxy. The discontinuity in the sky
brightness at these times is expected since the telescope is pointed
in a different direction. We refer to each of these sections of the
night as ``observing blocks''.

\begin{figure*}[t]
\begin{center}
\epsscale{1.0}
\plotone{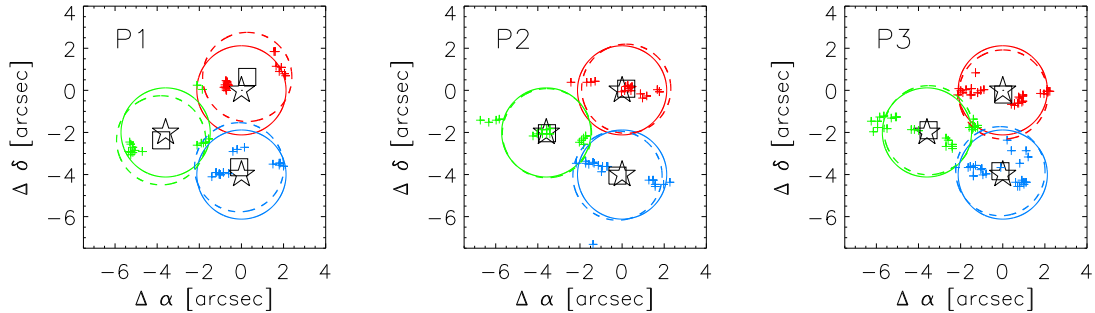}
\caption{Attempted and actual relative positions of the three sets of
  dithered exposures for the three pointings obtained on
  NGC 628. Stars and solid circles mark the attempted fiducial
  positions for dithers 1, 2, and 3 (red, green, and blue
  respectively). Crosses mark the actual position at which each
  exposure was obtained. Open squares and dashed color circles
  show the average fiber position of the actual observations.}
\label{fig-6}
\end{center}
\end{figure*}

Within each individual observing block, we fit the temporal evolution
of the sky brightness in each wavelength bin using a
cubic spline (color solid curves in Figure \ref{fig-5}). The best-fit splines allow
us to evaluate the overall spectral shape of the sky emission at any
arbitrary UT time within an observing block. This allows us
to calculate a wavelength dependent normalization curve by which the
two bracketing sky frames must be multiplied in order to reconstruct
the sky spectrum at the exact UT time of the science exposure. 
Since the normalization factors computed in the 500 wavelength
bins are inherently noisy, we fit them as a function of wavelength
using a fifth order polynomial. This allows us to multiply the sky
frames by a smooth function of wavelength containing the correction
without introducing further noise into the observed sky spectrum. 
We apply this procedure to correct the before and after
sky frames to the UT time of each science frame, and then
average the two corrected frames to create a single sky background
file associated with each science exposure. Sky subtraction is performed by VACCINE
using this composite corrected sky frame. 

Once this final sky frame is available, the method used by VACCINE to
subtract the sky spectrum is analog to the bspline algorithm used
to remove the solar spectrum from the twilight flats (see \S4.3). Briefly,
for each fiber in the science exposure the sky spectra of a set of
neighboring fibers in the background frame is simultaneously fit using
a bspline and is then subtracted from the science fiber spectrum
after scaling for the difference in exposure time between science and
sky frames. As mentioned above, this procedure
greatly benefits from the sub-pixel sampling of the sky obtained by
combining information from different fibers having slightly different wavelength
solutions. 

The quality of the sky subtraction in VENGA is excellent. We typically
see sky subtraction residuals that are fully consistent with Poisson
plus read-noise uncertainties. Larger systematic residuals usually appear at
the wavelengths of the 4 brightest sky emission lines in our wavelength
range.  These resiudals are caused by the fast time variability of these
spectral features, which is independent of the variability of the sky
continuum and broad emission bands which dominate our
corrections. These bright sky lines are masked during the analysis of
the science data. Rarely (less than 4\% of the frames in the case of
NGC 628), when
observations are taken under extremely bad observing conditions
(usually combinations of clouds and moon, or clouds and proximity to
twilight), obvious residuals in the sky subtraction can be
observed in the sky-subtracted science exposures. We reject these
frames from the dataset.

\subsection{Spectrophotometric Flux Calibration}

We use observations of spectrophotometric standard stars from
\cite{massey88} and \cite{oke90} to calibrate the VENGA spectra. The
method used to construct sensitivity curves from the IFU
observations of standard stars is described in \cite{blanc09}. The
only difference with the approach taken in that previous work is that
here standard stars are only used to perform the relative flux
calibration as a function of wavelength. The absolute scale of the
calibration comes from a comparison to broad-band optical images of
the VENGA galaxies. The absolute calibration is described in the next section. 

We calculate sensitivity curves for all standard stars taken during
each observing run. These curves are normalized to a common scale and
averaged to create a master sensitivity curve for each
run. All science frames obtained during each observing run are
multiplied by these curves to convert them to units of \mbox{erg s$^{-1}$
cm$^{-2}$ \AA$^{-1}$}. Since the atmospheric transparency at the time
of the science and standard star observations is unknown we need
the absolute scaling described in the following section. 
The error maps for each frame are scaled accordingly to the data. By
looking at the dispersion between different
sensitivity curves within each month, we estimate a typical
uncertainty in the relative flux calibration of $\sim8$\%.

\subsection{Astrometry and Absolute Flux Calibration}

The pointing of VIRUS-P is done using an offset guider camera
which images a $4.5'\times4.5'$ field $\sim 9'$ north of the IFU
science field. An astrometric calibration between the two
fields allows the observer to point the IFU by putting a guide star at
specific pixel coordinates on the guider detector. \cite{adams11a}
found systematic offsets of the order of $1''$ between the same pointings made
over different observing runs. These offsets are also present in our
data and must be taken into account before combining multiple frames. 
In order to accurately recover the astrometry of the VENGA science
observations we register reconstructed broad-band images of the
galaxies made from the IFU data to archival broad-band images. These
comparisons are also used to calibrate the spectra in terms of
absolute flux. For the NGC 628 blue and red IFU data we use the
SDSS-III\footnote{SDSS-III is managed by the Astrophysical Research
  Consortium for the Participating Institutions of the SDSS-III
  Collaboration including the University of Arizona, the Brazilian
  Participation Group, Brookhaven National Laboratory, University of
  Cambridge, Carnegie Mellon University, University of Florida, the
  French Participation Group, the German Participation Group, Harvard
  University, the Instituto de Astrofisica de Canarias, the Michigan
  State/Notre Dame/JINA Participation Group, Johns Hopkins University,
  Lawrence Berkeley National Laboratory, Max Planck Institute for
  Astrophysics, New Mexico State University, New York University, Ohio
  State University, Pennsylvania State University, University of
  Portsmouth, Princeton University, the Spanish Participation Group,
  University of Tokyo, University of Utah, Vanderbilt University,
  University of Virginia, University of Washington, and Yale
  University. }
DR8 $g$-band and $r$-band
mosaics\footnote{http://data.sdss3.org/mosaics}.

For every science exposure we integrate the
spectrum of each fiber over the corresponding SDSS transmission curve
($g$ or $r$ for blue and red setup data respectively)
in order to measure the monochromatic flux at the effective wavelength of the
broad-band filter. Simultaneously we convolve the SDSS image of the
galaxy  with a gaussian kernel to match the PSF to the seeing
under which the VIRUS-P data was taken and we perform aperture
photometry at the fiducial position of each fiber. We use circular
apertures that match the fiber size. After correcting the
\mbox{VIRUS-P} fluxes for atmospheric extinction at the airmass of the
observations we compare them to the SDSS fluxes and fit them using
the following expression:

\begin{equation}
f_{SDSS}=A\times f_{VP}+B
\end{equation}

\noindent
where $A$ is a normalization factor recovering the absolute flux
scale and $B$ recovers any residual background left from the sky
subtraction process. A perfect background subtraction in both the SDSS
image and the VIRUS-P spectra should translate in $B=0$. 

We perturb the fiducial astrometry of the VIRUS-P pointing by
applying offsets in both right ascension and declination in order to
minimize the $\chi^2$ of the fit. This registering process provides
corrected astrometry for each science frame. Given the wealth of
spatial information encoded in the relative brightness of hundreds of 
VIRUS-P fibers, the registering is very accurate and has a
typical uncertainty of $\sim0.1''$. The value of $A$ at the registered
position provides the absolute scale for the flux calibration and the
science frame is multiplied by this value. This ties our flux
calibration to that of SDSS which has a zero-point uncertainty of 2\%
in the bands used here. For the 195 individual frames used to construct the
NGC 628 data-cube, we
measure a mean $\langle B \rangle=-2\times 10^{-18}$ erg s$^{-1}$ cm
$^{-2}$ \AA$^{-1}$ with a standard deviation of $\sigma_{B}=4\times10^{-18}$
erg s$^{-1}$ cm $^{-2}$ \AA$^{-1}$. This level of sky subtraction residuals
correspond to less than 2\% of the median continuum level in the
data and this is only an upper limit since the residuals have some
contribution from the error in the SDSS sky subtraction.

Figure \ref{fig-6} shows the astrometric offsets we measure for the
three pointings on NGC 628 with respect to the
fiducial dithering pattern. For each pointing the fiducial positions
of a fiber on dithers 1, 2, and 3 are marked by black stars and solid red, green, and
blue circles respectively. Color crosses mark the actual positions at
which independent science exposures were obtained. These positions are
measured using the registration method described above. The black squares
and dashed color circles show the average position for all exposures
in each dither.

Overall the VIRUS-P acquisition accuracy is good compared
to the fiber size. We observe systematic offsets from the fiducial
dithering pattern in the range $0.1''$-$4.0''$, with a mean of
$1.7''$. This is in good agreement with the $1.8"$ accuracy found by
\cite{adams11a}. As mentioned above, once the VENGA frames are registered
to the broad-band images and these offsets are corrected we estimate
an astrometric uncertainty of $0.1''$.

When registering the data and building the final VENGA data-cubes (next
section) we ignore the effects of atmospheric differential refraction (ADR). At a
typical observation airmass of $\chi=1.2$ we expect less than $1''$ of ADR from the
blue to the red end of our data. Given the $4.2''$ fiber size
of VIRUS-P and the $5.6''$ FWHM PSF of the final data-cubes (\S4.7), we
decide not to apply an ADR correction. 

To evaluate the quality of our spectrophotometric calibration we
compare the VENGA spectrum of NGC 628 to the $20"\times20"$ long-slit driftscan
spectrum obtained by \cite{moustakas06} as part of the ancillary data
for the Spitzer Infrared Nearby Galaxies Survey \citep[SINGS,
][]{kennicutt03}. To do so, we use the final combined VENGA data-cube
described in the following section. We integrate the IFU spectra over
the same $20"\times20"$ region sampled by the SINGS long-slit
driftscan.

\begin{figure}[t]
\begin{center}
\epsscale{1.2}
\plotone{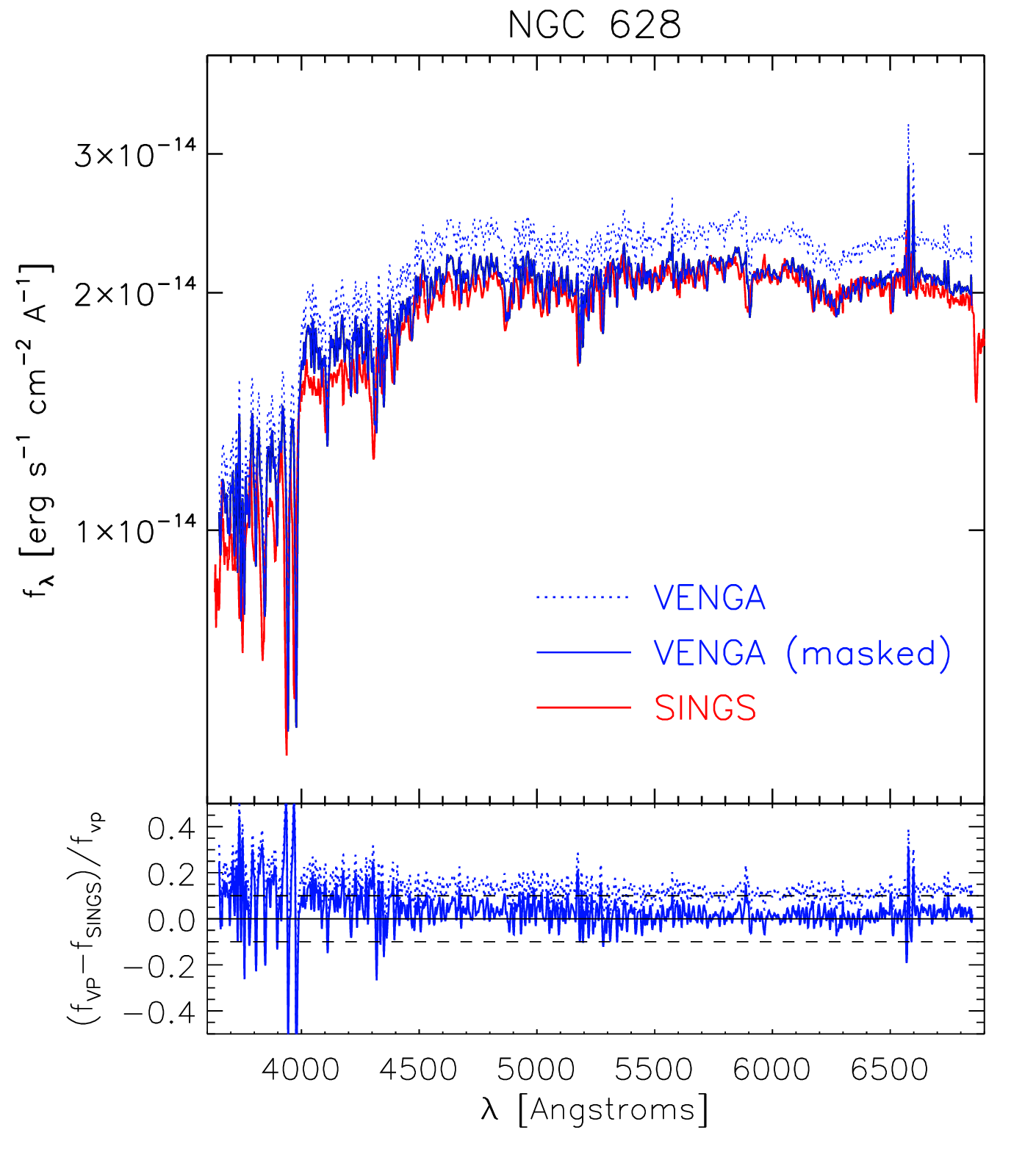}
\plotone{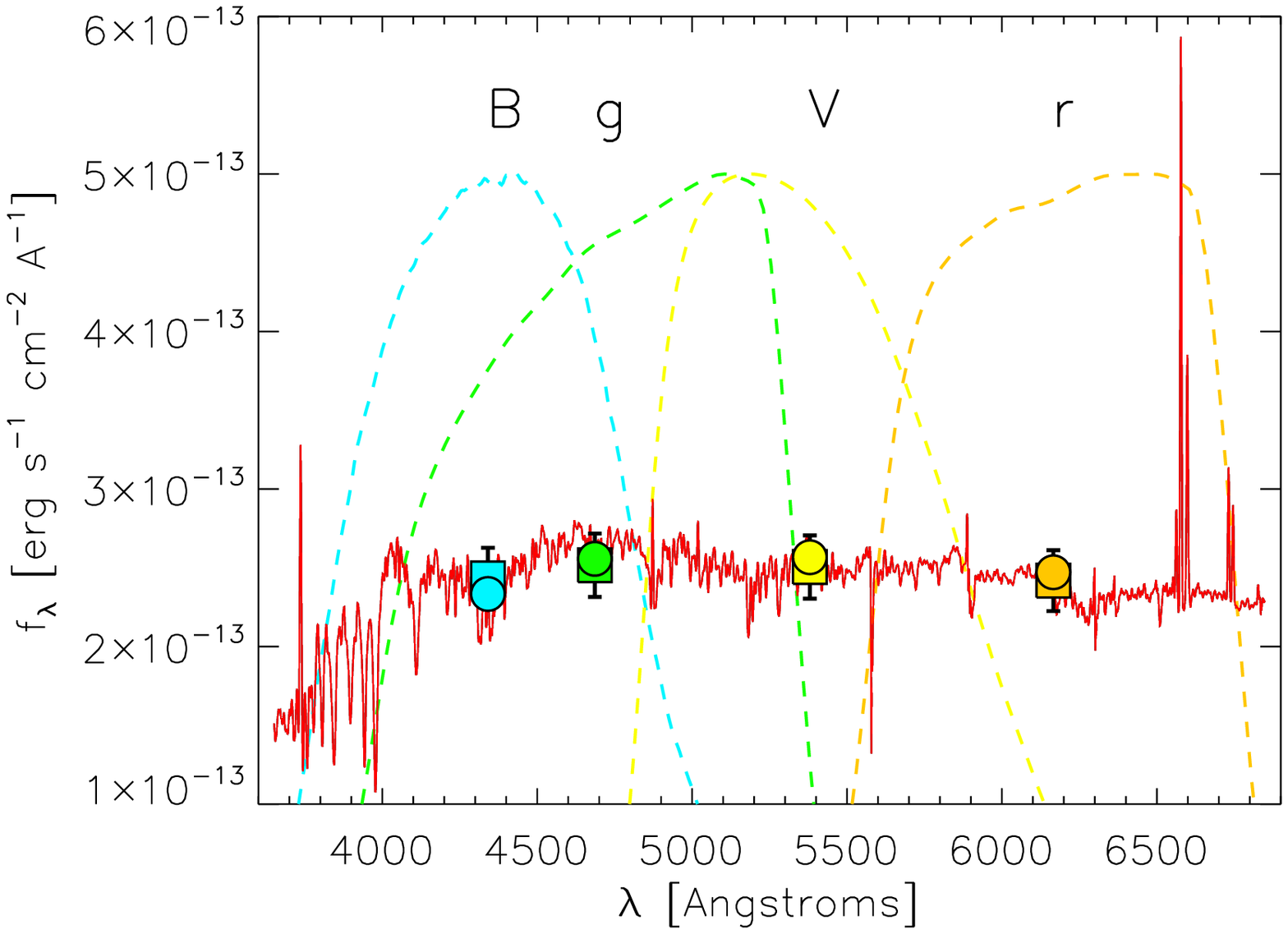}
\caption{{\it Top panel:} Comparison between the SINGS driftscan spectrum of the
  central $20'' \times 20''$ of NGC 628 \citep[red, ][]{moustakas06} and the VENGA
  IFU spectrum (dotted blue) integrated over the same region. 
The solid blue curve shows the VENGA spectrum after masking
  a foreground star which contaminates the measurement. Also shown are
  the residuals between the SINGS and VENGA spectra. Horizontal dashed
  lines mark a $\pm$10\% deviation. {\it Bottom panel:} Comparison
  between the monochromatic broad-band fluxes measured from the
  integrated VENGA spectrum over the whole datacube (squares) and the SDSS $g$ and
  $r$ and SINGS $B$ and $V$ band images (green, orange, light blue
  and yellow circles respectively). Error bars
  indicate the 8\% uncertainty in the relative flux calibration. Also
  shown for reference are the integrated VENGA spectrum (red) and the
broad-band transmission curves (dashed lines).} 
\label{fig-7}
\end{center}
\end{figure}

The comparison is shown in top panel Figure \ref{fig-7}. 
We observe an 18\% offset between the two spectra. This offset is
produced by contamination from the wings of a bright foreground star
which falls slightly outside the SINGS driftscan box, but which
significantly contributes to the flux inside the box at the $5.6''$
resolution of the VENGA data-cube. Masking the contaminated spaxels (see \S6.1) brings the
VENGA and SINGS spectra to agreement. Except for regions on top of strong emission
and absorption lines, where residuals arising from differences in
spectral resolution dominate, both spectra show less than 5\%
deviations from each other over 80\% of the overlap spectral range,
rising towards the 10\% level at the blue end of the spectrum.
This is well within the 10\%-30\% uncertainty in the absolute
calibration and the 3\%-4\% uncertainty in relative flux calibration
of the SINGS spectrum as estimated in \cite{moustakas06}, and the 8\%
relative flux error in the VENGA data (derived from the
dispersion between different sensitivity curves in \S4.5).

To further test the quality of the VENGA flux calibration we
compute broad-band monochromatic fluxes from the integrated spectrum over the
full VENGA datacube and compare them to the flux measured over the
same area in both the SDSS $g$ and $r$-band
images, and the SINGS $B$ and $V$-band optical ancillary
images\footnote{http://irsa.ipac.caltech.edu/data/SPITZER/SINGS/}. The
SINGS images are independent from our flux calibration. As shown in
the bottom panel of Figure \ref{fig-7} the VENGA data
agrees with the flux measured in all four images within the quoted
uncertainties. From these tests we conclude that the accuracy of the VENGA
absolute and relative flux calibrations is good.

\subsection{Final Combined Data-cubes and RSS Files}

At this stage in the data reduction process we have a reduced, sky
subtracted, wavelength and flux calibrated 2D spectrum of each
fiber (and its associated 2D error spectrum) in every individual
science exposure. By 2D spectrum we mean a non-collapsed spectrum of
the fiber PSF on the detector, which contains no spatial information
about the source, given the fact that optical fibers scramble the
input light. Also, the position of each fiber on the sky is known to
$0.1''$ accuracy thanks to the registering process described in the
last section. 

Here we describe the methods used to
extract and combine data from different exposures into a final
data-cube containing a 1D spectrum at each position on the sky.
We also describe the construction of our final row-stacked spectra
(RSS\footnote{RSS files, in which every row contains the spectrum of a
  spaxel, should be distinguished from data-cubes, as
they are two dimensional instead of three dimensional representations
of IFU data. They must be accompanied by a position table for all
spaxels in order to allow for a geometrical reconstruction of the data.}) files. These
multi-extension FITS\footnote{Flexible Image Transport System} files
are described below and contain the spectrum and error spectrum, as
well as the J2000 equatorial coordinates for each spaxel in the
data-cube. Also contained in the RSS file is information regarding the
central wavelength and spectral instrumental resolution for every
pixel in the spectra.

Data from different nights and observing runs have independent
wavelength solutions. Therefore, before combining we need to re-sample
all the spectra to a common wavelength grid. In VENGA we produce two
versions of the same data-cube, one with a regularly spaced linear
sampling of wavelength and another with a regular logarithmic sampling
(i.e. spaced regularly in velocity space). The {\it linear} version has
pixels spaced by 1.1\AA\ (similar to the average dispersion in the
original data), while the {\it logarithmic} data-cubes have pixels
that are spaced by $(\Delta \lambda/\lambda)c=60$ km s$^{-1}$. 
We work with spectra in flux density units so total flux is
properly conserved when re-sampling. 

The reason behind producing
two versions of the data is that while most users will be
interested in using the {\it linear} version for many applications, the
spectral fitting software used in the following section to extract
stellar and gas kinematics, as well as emission line fluxes, requires
input spectra that is regularly sampled in velocity space. Instead of
interpolating the spectra to a linear grid for the effects of
combining, and later re-interpolating the combined spectra to a
logarithmic scale, we do both re-samplings directly from the original
data. In this way we avoid the effects of S/N degradation associated
with extra interpolations.

\begin{figure}[t]
\begin{center}
\epsscale{1.2}
\plotone{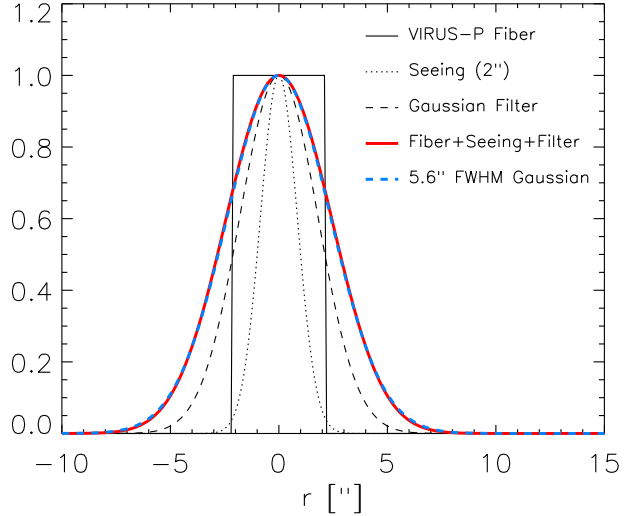}
\caption{The final PSF of the VENGA data-cubes given by the
convolution of the fiber profile (solid black line) with the seeing
(dotted black line) and the Gaussian spatial filter used to combine
the individual exposures (dashed black line) is shown in red. The PSF
is well described by a Gaussian PSF with $5.6"$ FWHM (dashed blue
line), and is largely independent of the seeing.} 
\label{fig-8}
\end{center}
\end{figure}

For each fiber in each science frame, after re-sampling the 2D spectra
we have 5 pixels at any given
wavelength which provide a measurement (with its associated error) of the flux
density at a certain position on the sky. We then define a regularly
spaced spatial grid of spaxels. The grid has a spaxel scale of $2''$
chosen to roughly Nyquist sample the final data-cube PSF. For every spaxel in
the grid we compute the final spectrum by combining the spectra of
surrounding fibers using a Gaussian spatial filter and adopting an
inverse variance weighting scheme. Therefore, the spectrum of each
spaxel in the data-cube is given by

\begin{equation}
f_{\lambda} (\lambda)=\frac{\sum\limits_{i,j,k} w_{i,j,k}(\lambda) f_{\lambda}^{i,j,k}(\lambda)}{\sum\limits_{i,j,k} w_{i,j,k}(\lambda)}
\end{equation}

where $f_{\lambda}^{i,j,k}(\lambda)$ is the flux density of pixel $k$
($k=1-5$) at wavelength $\lambda$ of fiber $j$ in
frame $i$, and the weights $w_{i,j,k}$ are given by

\begin{equation}
w_{i,j,k}(\lambda)=\frac{1}{(\sigma_{i,j,k}(\lambda))^2} \times e^{\frac{-d_{i,j}^2}{2\sigma_{g}^2}}
\end{equation}

where $\sigma_{i,j,k}(\lambda)$ is the error in the flux, $d_{i,j}$ is the
distance in arcseconds between the fiber center and
the spaxel in question, and $\sigma_g$ is the width of
the spatial Gaussian filter used to combine the data. We use a value of
$\sigma_g=4.24/2.355$ which matches the fiber size to the FWHM of the
Gaussian filter. The sum in Equation 2 is
performed over all fibers within a maximum distance of the spaxel of
interest. This distance cut is given by a 99\% drop in the Gaussian
filter weight. During the combination process we apply 3$\sigma$ clipping rejection
to remove any cosmic rays that were not masked by LA-Cosmic in \S4.1. The final
spatial PSF of the data-cube is given by the convolution of the top-hat $4.24"$
diameter fiber profile with the seeing ($2.0"$ FWHM for the NGC 628
data) and the $4.24"$ FWHM Gaussian spatial filter used to combine the
data. Figure \ref{fig-8} presents these three components and the
product of their convolution (red solid line). The final PSF is well
described by a $5.6"$ FWHM Gaussian and is largely independent of the
seeing. Direct measurement of the FWHM of bright foreground
stars in the NGC 628 data-cube empirically confirms the value of $5.6"$. 

As mentioned in \S4.2, the instrumental spectral resolution at
different wavelengths, for each fiber, is extracted from the arc
lamps. To create a spectral resolution map for the data-cube, we combine
the master arc lamp frames associated with each individual science frame
in the same way as the science data (i.e. using the same
weights). We then use the method described in \S4.2 to create a map of
the instrumental spectral resolution from this combined arc.

The final VENGA data products for each galaxy are a data-cube
(used mostly for visualization purposes) and a RSS file (used for
spectral fitting and analysis). The data-cube consists of a
multi-extension fits file, in which each extension corresponds to an
image of the galaxy at a given wavelength (sampled at either linear or
logarithmic steps as described above). The data-cubes are in units of
flux density (erg s$^{-1}$ cm$^{-2}$ \AA$^{-1}.$).

The corresponding linear and logarithmically sampled VENGA RSS files are stored as multi-extension FITS files,
and contain the following information in their different extensions:

\begin{figure*}
\begin{center}
\epsscale{1.1}
\plotone{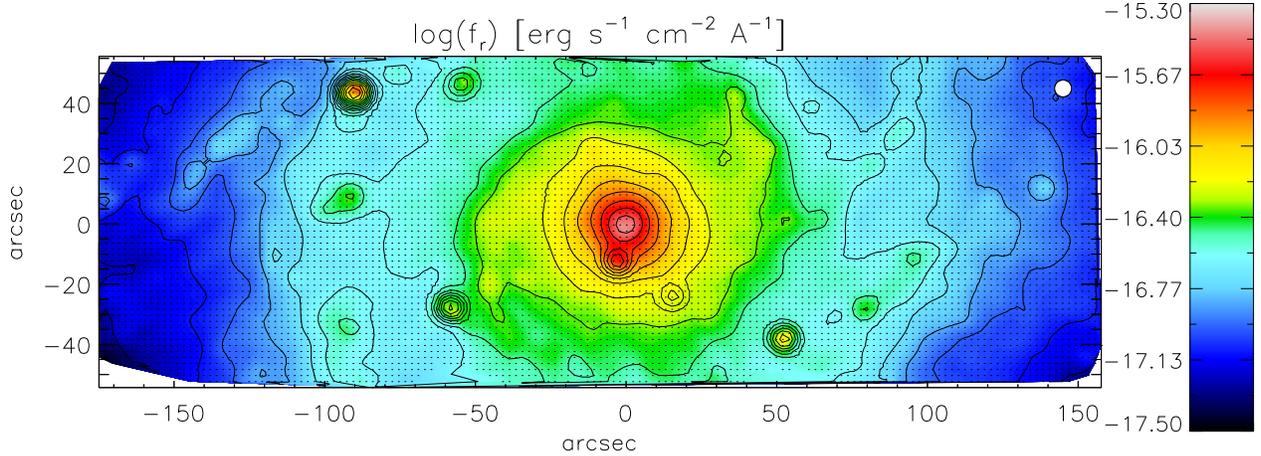}
\caption{Map of the $r$-band flux reconstructed from the VENGA spectral
  data-cube of NGC 628. Black contours show isophotes in the SDSS
  $r$-band image of the galaxy (PSF matched to the $5.6''$ FWHM PSF of the
  VENGA data which is shown by the white circle in the top right
  corner). Black dots mark the position of each spaxel in the
  data-cube. This and all maps presented in this work were constructed using the
PLOT\_VELFIELD IDL routine written by Michele
Cappellari (http://www-astro.physics.ox.ac.uk/$\sim$mxc/idl/).}
\label{fig-9}
\end{center}
\end{figure*}

\begin{enumerate}
\item Flux density spectrum for each spaxel in units of erg
  s$^{-1}$ cm$^{-2}$ \AA$^{-1}.$
\item Error spectrum for each spaxel in units of erg
  s$^{-1}$ cm$^{-2}$ \AA$^{-1}.$
\item Central wavelength of all pixels in the spectrum of each spaxel in units of \AA.
\item Right ascension and declination of each spaxel in units of
  decimal degrees.
\item Instrumental spectral resolution (FWHM) as a function of
  wavelength for each spaxel in units of \AA.
\end{enumerate}

The final VENGA data-cube of NGC 628 covers an area of
$5.2'\times1.7'$ and samples the disk of this galaxy out to 8 kpc ($0.6R_{25}$). Figure
\ref{fig-9} shows an \mbox{$r$-band} image of NGC 628 reconstructed from the
final VENGA data-cube. All maps presented in this paper have north
pointing up and east pointing left. Black dots mark the position of
each spaxel in the data-cube and the white circle on the top
right corner shows the PSF FWHM. The final combined spectra was
integrated over the SDSS \mbox{$r$-band} transmission curve to create
this map. For comparison, isophotal contours from the SDSS $r$-band
image (with the PSF degraded to $5.6''$) are overlaid in black.
The similarity between the two surface brightness distributions
confirms that the flux calibration and astrometric correction of
the individual science frames has been
done properly. The spectra in the final data-cube of NGC 628 has a
median $S/N=117$ in the continuum per spectral resolution element (FWHM). In the
central parts of the galaxy we typically have $S/N>300$, while the
spectra in the faintest regions has $S/N\sim12$. In terms of surface
brightness depth, the NGC 628 data reaches a 5$\sigma$ detection limit
for continuum at a $V$-band surface brightness of 25.0 mag
arcsec$^{-2}$ (AB). We remind the reader that the data on NGC 628 is
about a factor of two deeper than for the average VENGA target.

\section{Spectral Analysis Pipeline}

In order to extract emission line fluxes, gas and stellar kinematics,
and information about the stellar populations present in different parts
of the galaxies, we fit the VENGA spectra using a linear combination of
templates convolved with a LOSVD plus a set of
Gaussian emission line profiles. To do the fitting we use the pPXF
\citep{cappellari04} and GANDALF \citep{sarzi06} IDL routines
developed for this purpose by the SAURON team. In this section we
describe the fitting process and present an example fit in the NGC
628 data-cube. We also discuss how the uncertainty in the fitted
parameters is calculated. In this paper we fit the stellar continuum using
empirical stellar templates (\S5.1). The same type of fitting can be
performed using stellar population synthesis (SPS) templates to extract
information about the stellar populations present in different regions
within the galaxies. The SPS fitting of the VENGA galaxies will be
presented in a future publication. 
Before fitting the spectra, the data is corrected for Galactic
extinction adopting a Milky Way (MW) extinction law as parametrized by
\cite{pei92}, and extinction values from the maps of
\cite{schlegel98} ($E(B-V)_{MW}=0.07$ mag or $A_{V,MW}=0.23$ mag for NGC 628).

\subsection{Stellar Kinematics}

We mask the spectrum of each spaxel around regions affected by
sky subtraction residuals due to bright sky lines, regions
potentially affected by the nebular emission lines listed in Table
\ref{tbl-6}, and the edges of each of the individual instrumental
setups (red and blue). We then fit for the stellar line-of-sight velocity
($v_{*}$) and velocity dispersion ($\sigma_{*}$) with the
pPXF software, which uses the ``penalized pixel'' fitting technique
\citep{cappellari04} to fit the spectrum as a linear combination of templates
convolved with an LOSVD. Although the software uses a Gauss-Hermite polynomial
LOSVD, and allows for the fitting of high order terms
($h_3$,$h_4$), for the data presented here we only fit for the first
two moments $v_*$ and $\sigma_*$.

The logarithmically sampled RSS files are used as input for
pPXF, which requires the spectra to be regularly sampled in
velocity space. We use the MILES stellar library version 9.1
\citep{sanchez-blazquez06, falcon-barroso11} as a source of empirical templates. A
subset of 48 stars spanning a wide range in spectral types
(O through M), luminosity classes (I through V), and metallicities
($-2<$[Fe/H]$<1.5$) is used. Also included in the subset are horizontal branch and
asymptotic giant branch (AGB) stars.

Before fitting, the templates are re-sampled to the wavelength scale of the
data. We match the spectral resolution of all spaxels in the VENGA
data and the templates to the worst instrumental resolution at
any given wavelength in the data-cube by convolving the spectra with a
running Gaussian kernel with a wavelength dependent width. This
translates in a spectral resolution of $\sigma_{ins}=110$ km s$^{-1}$
in the red edge and $\sigma_{ins}=200$ km s$^{-1}$ in the blue end of
the spectrum. The median instrumental resolution across the convolved
data-cube is $\sigma_{ins}=150$ km s$^{-1}$. To convolve the templates
we assume the corrected MILES library intrinsic resolution of 2.54\AA\
\citep{beifiori11}. 

Close inspection of fits to very high signal-to-noise spaxels in the data-cube
($S/N \ge 100$) shows systematic residuals caused by template mismatch at the 1\%
level in the 5000\AA-6800\AA\ region, and rising up to 2\% towards the blue
end of the wavelength range. It is important to include this
systematic uncertainty floor during the fitting process. Particularly for
high $S/N$ spectra where it becomes the dominant source of discrepancy
between the best fit models and the data. We assume a
systematic uncertainty of 2\% at 3600 \AA, linearly decreasing
towards 1\% at 5000 \AA and staying flat at this level redwards. We
add this uncertainty in quadrature to the photometric errors before
fitting the spectra.

In order to account for the effect of dust extinction on the shape of
the continuum as well as systematic differences in the flux calibration of the data
and the templates, during the minimization we fit both additive and
multiplicative low order Legendre polynomials. The low order of
the polynomials prevents them from introducing features on small scales
of the order of the instrumental resolution. The polynomials only
match the large-scale ($\sim 100$\AA) shape of the continuum in the linear
combination of stellar templates and the data and they do not affect
the fitting of individual spectral features. 

We fit the spectrum of each spaxel individually and store the kinematic
parameters ($v_*$, $\sigma_*$). We limit the wavelength range to the Mg I $b$
absorption feature (5100-5300 \AA) for the purpose of extracting the
stellar velocity dispersion. We do this to avoid
a systematic overestimation of $\sigma_*$ caused by
the broad wings of Balmer absorption lines which are subject to
rotation and pressure broadening. As discussed below, this
effect is particularly significant in the spectra of star-forming
regions which have a large contribution from A type stars. We also
experimented with using other strong absorption features like Ca H+K
and the G-band, finding that Mg I $b$ provides a better agreement with
the published high precision measurement of the velocity dispersion
profile of NGC 628 by \cite{herrmann08} (see \S6.4) while the other
features always produced systematically higher values of
$\sigma_{*}$. Issues regarding the overestimation of $\sigma_{*}$ when
using Balmer lines, Ca H+K and the G-band have
been extensively discussed in the literature
\citep[e.g.][]{kormendy82, bernardi03, greene06}.
The instrumental resolution in the convolved spectra at the Mg I $b$ feature
is $\sigma_{\rm ins}=145$ km s$^{-1}$.

We also perform a second fit using the full wavelength range to
estimate $v_*$. Our simulations show that while the best fit
$\sigma_*$ is biased when using the full spectrum, the best fit $v_*$ is not and the
uncertainty in its value is reduced significantly by using
the full wavelength range. After fitting, we keep the LOSVD parameters fixed over
the next fitting iteration (\S5.3), in which we also fit for the emission
lines in the spectrum. Kinematic maps of NGC 628 are presented in \S6.

\subsection{Error Estimation for LOSVD Parameters}

To estimate the uncertainty in the best-fit LOSVD parameters ($v_*$
and $\sigma_*$) we fit a series of Monte Carlo realizations of
simulated spectra with different stellar velocity dispersion,
signal-to-noise ratio, and combination of stellar templates. To
construct the simulated synthetic spectra we start with the best-fit linear
combination of stellar templates for three different regions in NGC
628 chosen to sample an old stellar population with a spectrum
dominated by K giants (taken from the central part of the
galaxy), a bright star forming HII region on a spiral arm, and a region
showing a post-starburst spectrum dominated by A type stars. 
The three simulated spectra are convolved with Gaussian LOSVDs
having $\sigma_{\rm *,true}=$10, 25, 50, 100, 200, 300, and 500 km s$^{-1}$.
Gaussian noise is artificially added to the spectra creating 40 Monte Carlo
realizations of each at $S/N$ levels per resolution element (FWHM)
of 25, 50, 100, and 200.

We fit the synthetic spectra using the same methods used for the data
and described in the previous section. Even though the simulated
spectra do not contain emission lines we mask the same regions which
we censored when fitting the science data in order to match the
available spectral information. In an attempt to account for the
effects of template mismatch we remove
the templates used to create the synthetic spectra from the list of
templates used to fit them. The large number of templates used
($\sim50$) allows us to do this without the worry of a particular
stellar type not being represented in the remaining template subset.
We present results for two fits, one using the full available spectral range, and
one limited to the 5100-5300 \AA\ region around the Mg I $b$ feature.

\begin{figure}
\begin{center}
\epsscale{1.2}
\plotone{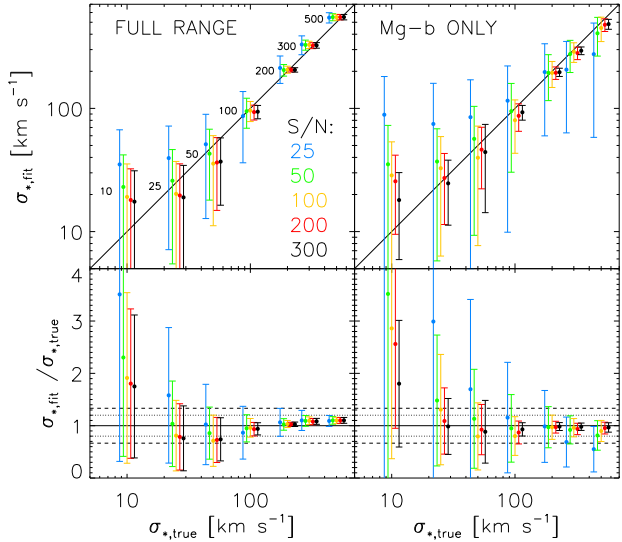}
\caption{{\it Top panel:} Comparison between the measured
  ($\sigma_{*,fit}$) and the input ($\sigma_{*,true}$) velocity
  dispersion from fits to Monte Carlo realizations of simulated
  spectra. Colors indicate the $S/N$ of the synthetic
  spectra. Groups of datapoints for different values of $\sigma_{*,true}$ are
  labeled in the upper left panel. Filled
  circles and error-bars indicate the mean and standard deviation of
  the recovered values from 40 Monte Carlo realizations. Points are
  shifted by small amounts along the horizontal axis for clarity. {\it Bottom
    panel:} Ratio between the recovered and input velocity
  dispersions. Dotted and dashed horizontal lines mark 20\%
  (5$\sigma$ measurement) and 33.3\% (3$\sigma$ measurement)
  deviations respectively}
\label{fig-10}
\end{center}
\end{figure}

\begin{figure}[t]
\begin{center}
\epsscale{1.2}
\plotone{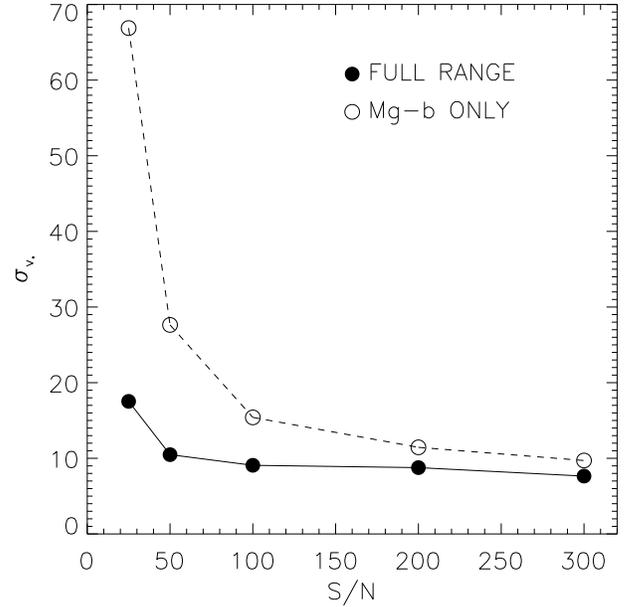}
\caption{Error (standard deviation) in the recovered stellar velocity for fits
to simulated spectra with $\sigma_{*,true}<100$km s$^{-1}$ as a function of
signal-to-noise ratio. Filled and open circles correspond to fits
using the full spectral range and fits limited to the 5100\AA-5300\AA\
window respectively. }
\label{fig-11}
\end{center}
\end{figure}

The results of the simulation are presented in Figures
\ref{fig-10} and \ref{fig-11}. In Figure \ref{fig-10} we compare the
input and recovered values for the velocity dispersion at different
$S/N$ ratios. In Figure \ref{fig-11} we present the error in the
recovered radial velocity as a function of $S/N$ for all the simulated
spectra with $\sigma_{*}<100$km s$^{-1}$ (typical of spiral galaxies).
Results for fits using both the full spectral range and the limited
5100\AA-5300\AA\ window are reported.

When using the full spectral range the recovered stellar velocity dispersions are
overestimated by $\sim 10$\% at $\sigma_{\rm *,true}>100$ km s$^{-1}$,
become underestimated by $\sim 30$\% below that value, and then become
largely overestimated for values of $\sigma_{\rm *,true}$ which are
about a factor of 10 smaller than the instrumental resolution. In the
case of the full spectral range fit this trends are almost independent
of the $S/N$ ratio. As can be seen in the right panel of Figure \ref{fig-10},
limiting the spectral window to the Mg I $b$ region removes most of
these systematic deviations at the cost of larger random
error-bars on the recovered $\sigma_{*}$ values. Systematic deviations
are still seen at low $S/N$ and low $\sigma_{\rm *,true}$. Balmer lines play a
significant role in introducing the systematic deviations seen in the
full spectral range case as is confirmed by the fact that the effect is stronger for the ``star-forming''
and ``post-starburst'' spectrum, than it is for the ``old stellar
population'' spectrum. No significant difference is seen on the
recovered velocity dispersions for the three different types of
spectra once the wavelength range is limited to the Mg I $b$ feature.

 An effect of limiting the spectral window is an
increase in the random error in the recovered
value of $\sigma_*$. This increase is expected because of the high EW of the
rejected Balmer absorption lines. The error in $\sigma_*$ for each
spaxel in the data-cube is computed by linear interpolation of the
measured errors in the simulation to the spaxel's $S/N$ and
$\sigma_*$.

The radial velocity is recovered with no systematic deviations. Considering
all simulated spectra with $\sigma_{\rm *,true}<100$ km s$^{-1}$ (typical of the disks of
spiral galaxies) and at all the $S/N$ ratios probed, we measure mean
offsets of $-3\pm11$ and $-2\pm15$ km s$^{-1}$ for the fits using the
full spectral range and the 5100\AA-5300\AA\ window respectively.
The error in the recovered radial velocity depends on $S/N$ as
shown in Figure \ref{fig-11} and it is reduced by about a factor of
two at the typical $S/N$ ratio of the data by using the full spectral range including the strong Balmer
lines. The improvement is even more dramatic at low signal-to-noise
($S/N<50$), where the Mg triplet starts to get lost in the noise
while the Balmer lines can still be easily seen in the spectra.
Errors in $v_*$ for each spaxel are computed from the $S/N$ by linear
interpolation of the simulation results (black line in Figure
\ref{fig-11}), and adding in quadrature the wavelength solution r.m.s
of 6 km s$^{-1}$(\S4.2).

\begin{figure*}
\begin{center}
\epsscale{1.0}
\plotone{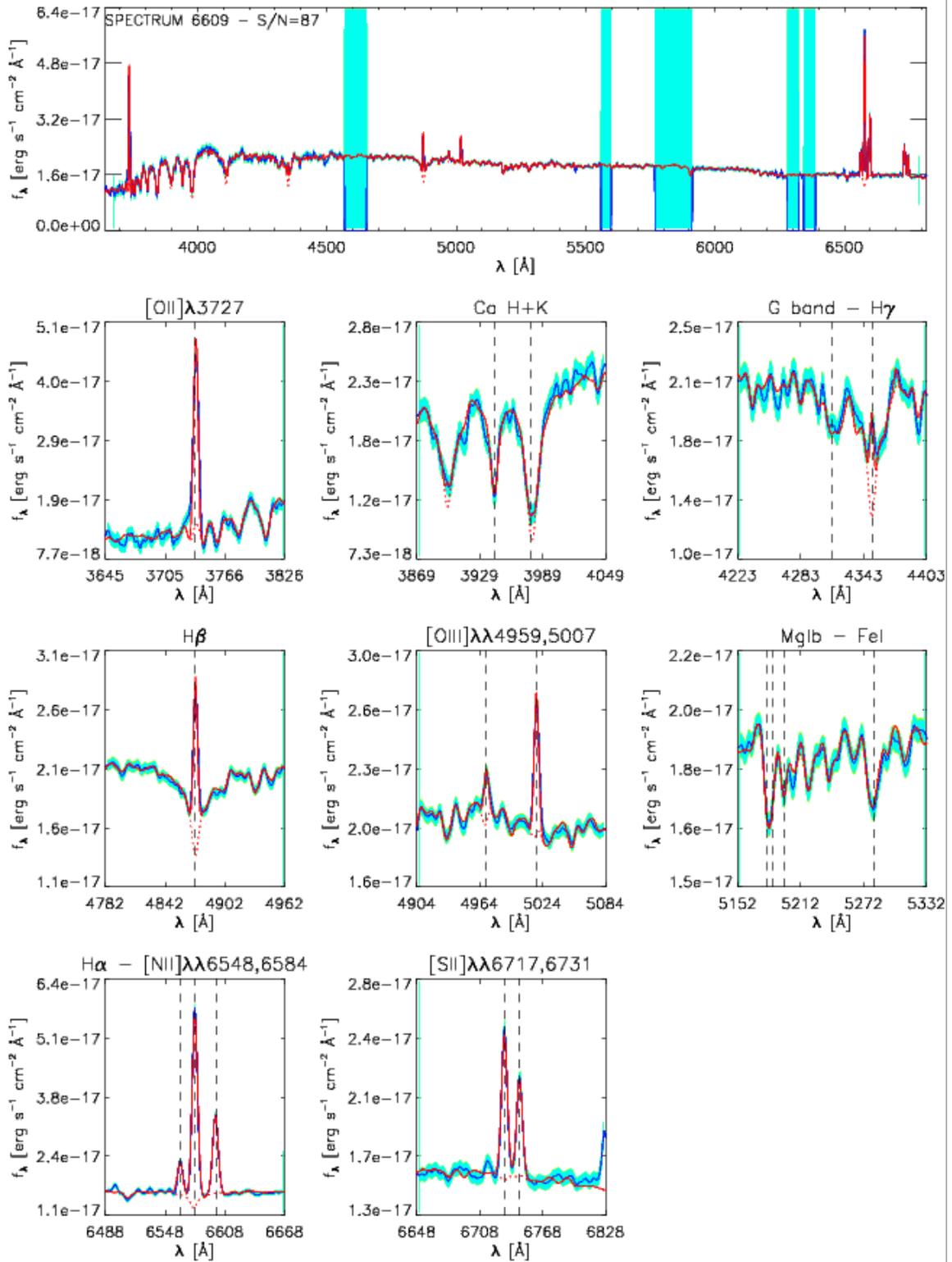}
\caption{Observed and best-fit spectrum of a randomly selected spaxel
  in the NGC 628 data-cube. The spectrum has a $S/N$ ratio of 87 in
  the continuum per spectral resolution element. The observed spectrum
  (solid blue) is shown with photometric errors (cyan envelope) and
total errors (photometric plus systematic, green envelope). Also shown
is the best-fit stellar plus emission line
spectrum (solid red), and the best-fit stellar spectrum alone (dotted red).
Vertical cyan bands represent regions masked around sky line
residuals and the ends of the blue and red setup spectra. Also shown
are zoomed in relevant spectral
features.}
\label{fig-12}
\end{center}
\end{figure*}

\subsection{Emission Line Fluxes and Ionized Gas Kinematics}

After measuring the LOSVD of each spaxel we use GANDALF
to fit the spectrum including the emission lines. GANDALF recomputes
the weights given to the different stellar templates at the same time of adding
Gaussian profiles to model the contribution from emission
lines. We attempt to fit all the transitions listed in Table \ref{tbl-6}.

During this fit the stellar LOSVD computed in the previous section is
held fixed but the emission lines velocity, velocity dispersion, and amplitude, as
well as the weights given to the stellar templates are free parameters. 
While in principle we could fit the kinematics of different lines independently, this
becomes very hard for faint transitions detected at low $S/N$. Therefore we
tie the kinematics of all emission lines to a common set of parameters
($v_{gas}$, $\sigma_{gas}$) during the fit. This ensures that the
kinematic parameters are dominantly constrained by the brightest
emission lines in the spectrum (typically H$\alpha$,
[OIII]$\lambda$5007, and [OII]$\lambda$3727). In this second iteration
we also use additive and multiplicative Legendre polynomials to match
the continuum shape.

Figure \ref{fig-12} presents the observed and \mbox{best-fit} spectrum of a
randomly selected spaxel in the data-cube. This spectrum provides a good
representation of the quality of our fits. The observed spectrum is shown
in blue, with photometric errors marked by the cyan envelope, and
total errors (photometric plus systematic) shown as a green envelope. The solid red
line is the best-fit stellar plus emission line
spectrum and the dotted red line shows the best-fit stellar
spectrum alone. The vertical cyan bands represent regions masked around sky line
residuals and the ends of the blue and red setup spectra. Also shown
are zoomed in windows around relevant spectral
features. Overall GANDALF produces fits to the VENGA spectra of
excellent quality.

\subsection{Errors in Emission Line Parameters and Kinematics}

GANDALF returns emission line parameter errors based on the covariance
matrix computed during the non-linear Levenberg-Marquardt fitting. 
Therefore, the quality of this error estimates depends
critically on the uncertainty in the observed spectrum being 
properly estimated. In particular for low EW emission lines detected in very
high $S/N$ spectra, for which the formal photometric errors are small,
the inclusion of the systematic uncertainties
associated with template mismatch (described in \S5.1) is crucial in 
order to obtain adequate emission line flux
errors. If this floor of systematic uncertainty is not included,
errors in the measured emission line parameters can be significantly
underestimated.

\begin{figure}
\begin{center}
\epsscale{1.1}
\plotone{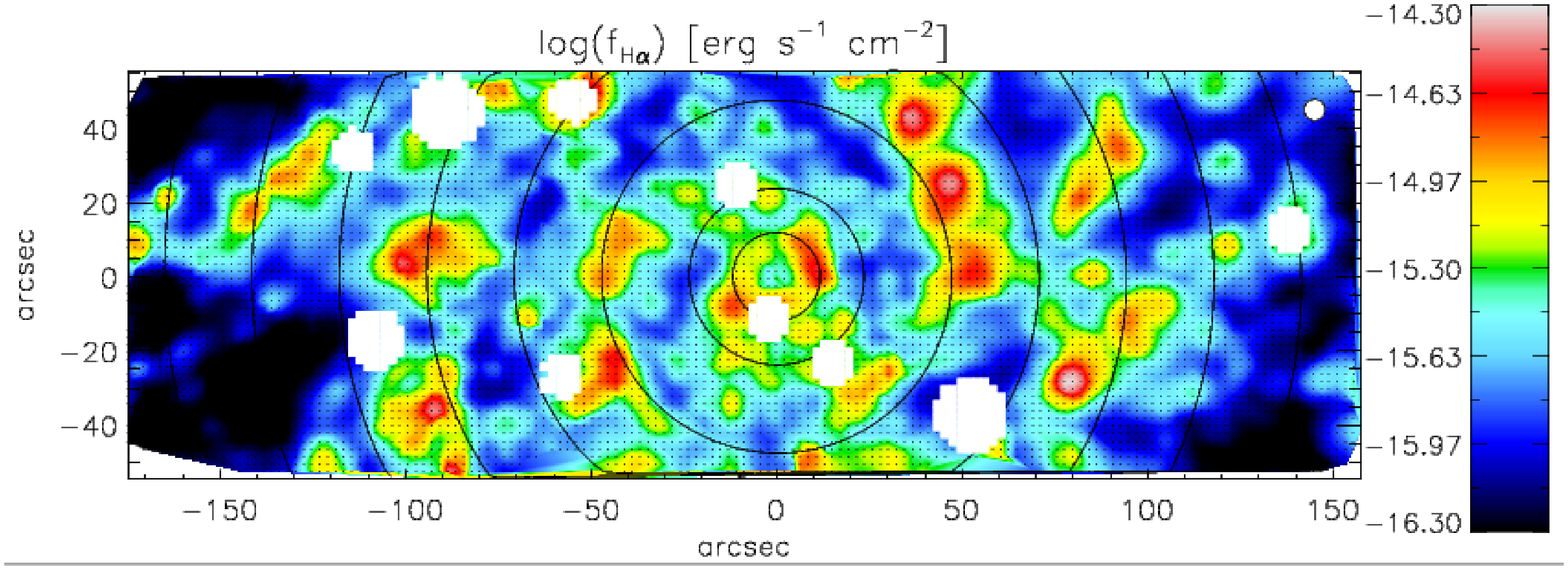}
\plotone{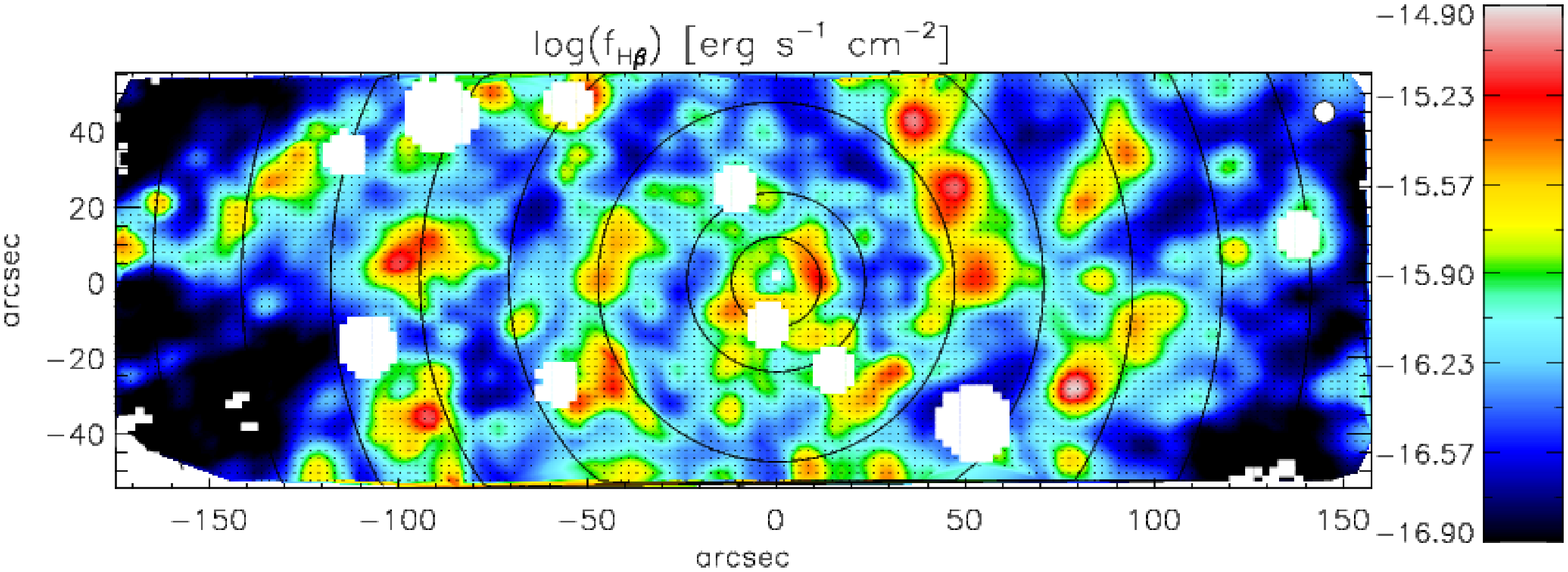}
\plotone{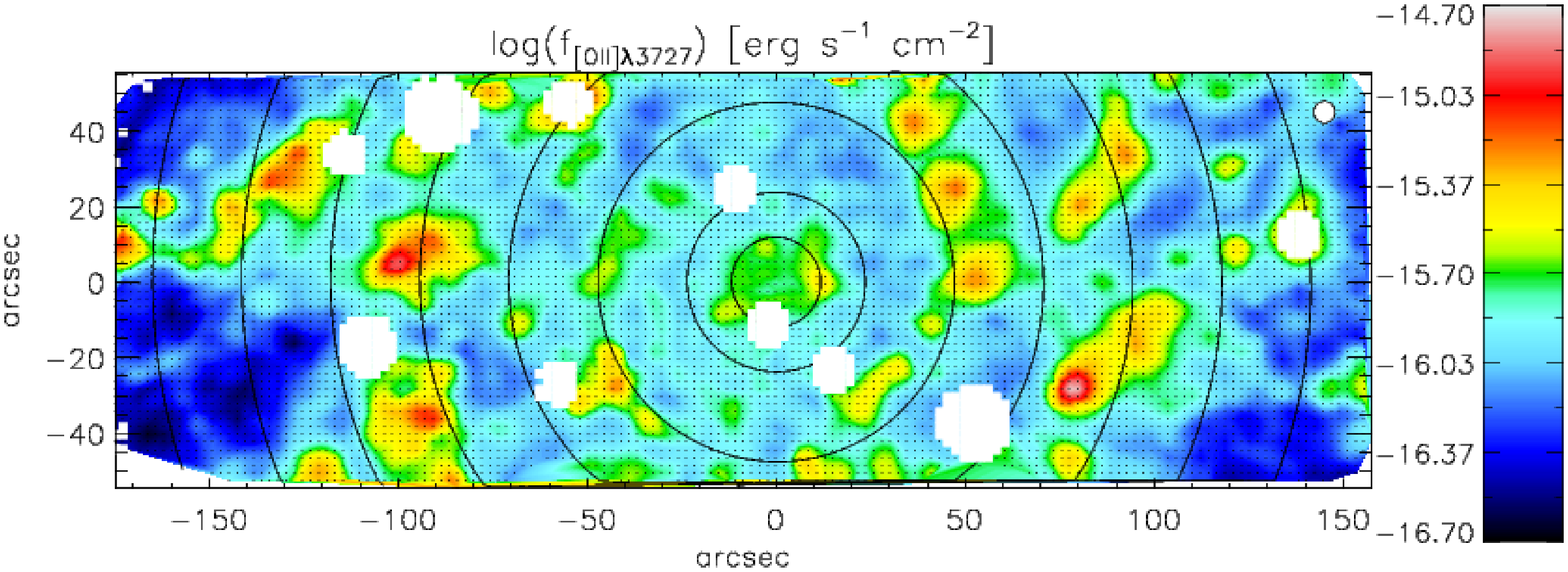}
\plotone{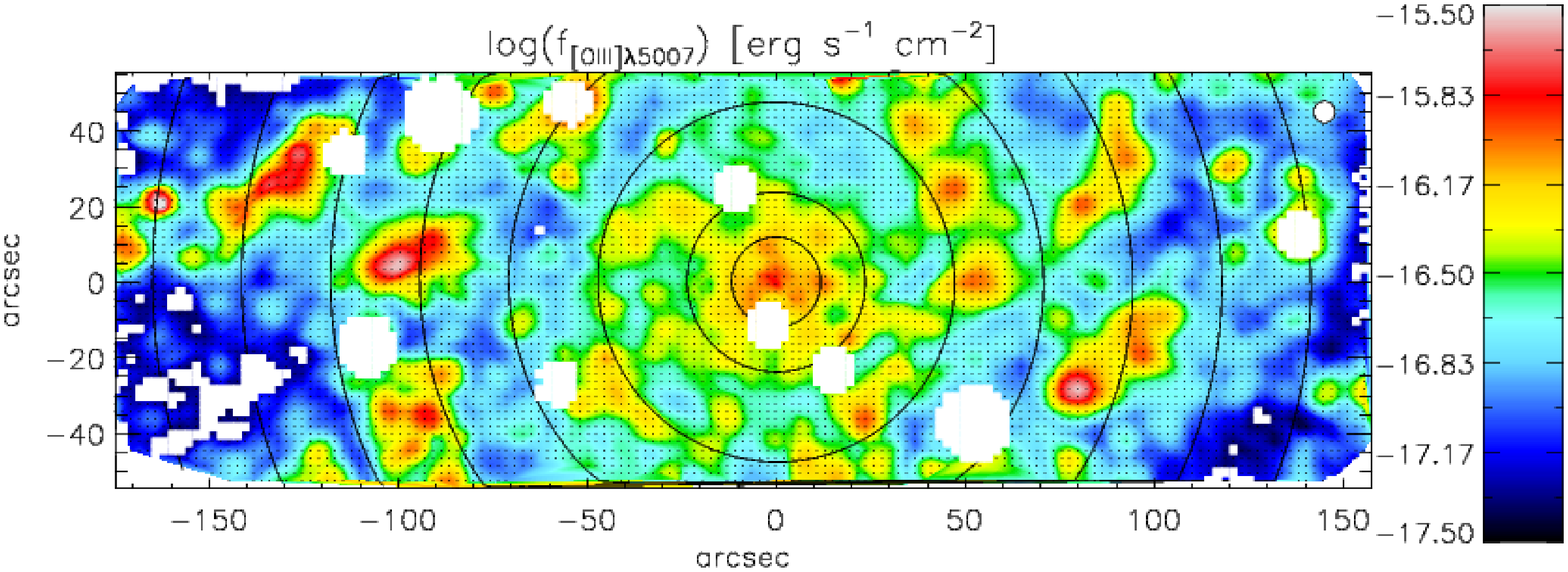}
\plotone{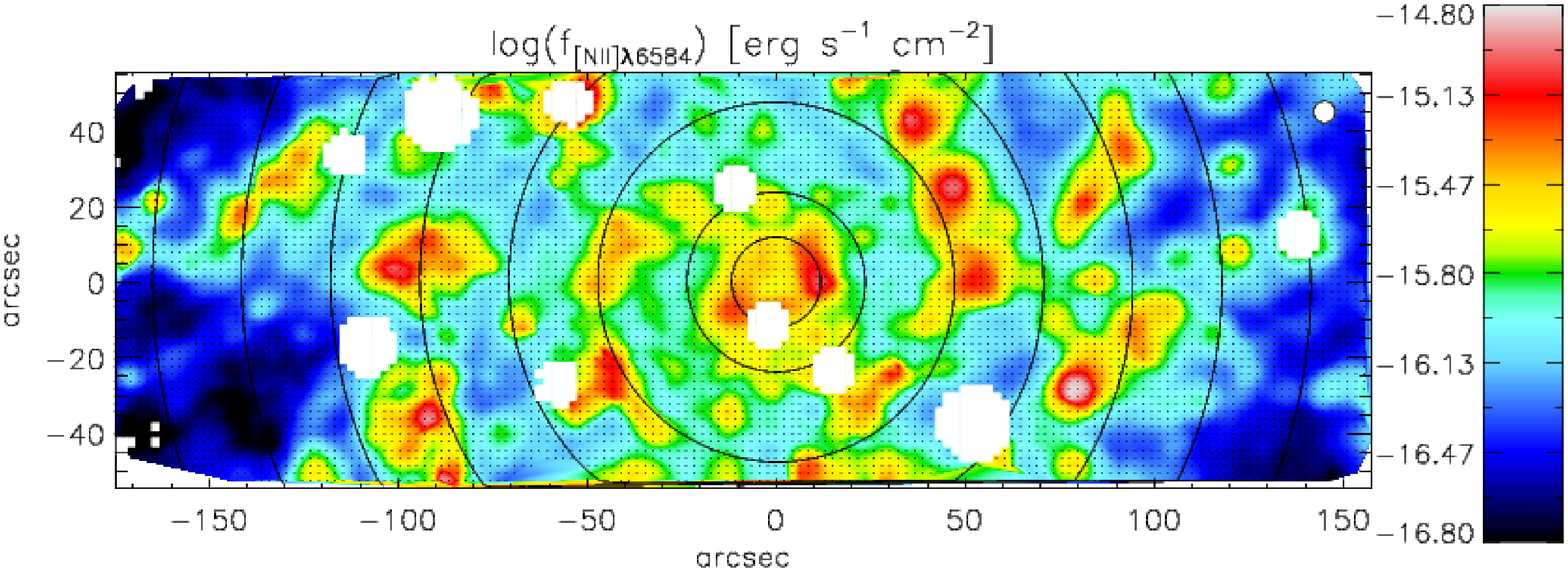}
\plotone{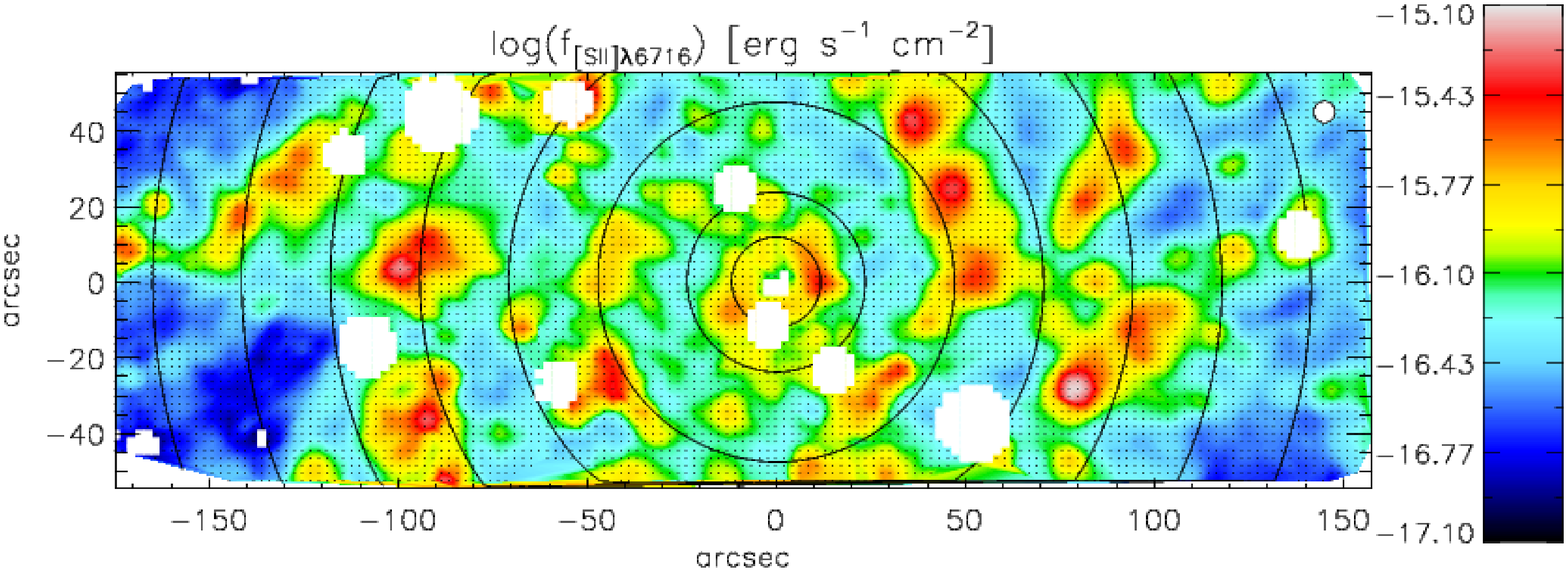}
\caption{Maps of NGC 628 in different emission lines. From top to
  bottom: H$\alpha$, H$\beta$,
[OII]$\lambda$3727, [OIII]$\lambda$5007, [NII]$\lambda$6584, and
[SII]$\lambda$6716. Black contours show lines of constant
galactocentric radii spaced by 1 kpc, with the exception of the
innermost contour at a radius of 0.5 kpc. Spaxels contaminated by
foreground MW stars (see \S6.1) are masked.}
\label{fig-13}
\end{center}
\end{figure}

Figure \ref{fig-13} shows maps of the flux of the H$\alpha$, H$\beta$,
[OII]$\lambda$3727, [OIII]$\lambda$5007, [NII]$\lambda$6584, and
[SII]$\lambda$6716 lines across \mbox{NGC 628}. We only show spaxels in which
the line is detected with a significance higher than 3$\sigma$. The
H$\alpha$ line is
detected at 5$\sigma$ over the full data-cube. 
The nebular emission clearly traces the two main spiral arms of the galaxy where ongoing star
formation gives rise to prominent HII regions. In the inter-arm
regions we still detect significant amounts of H$\alpha$ emission
although with a surface brightness that is one
to two orders of magnitude fainter than in the arms. A large fraction
of this emission arises from the diffuse ionized gas component of the
galaxy's ISM and is enhanced in low ionization ratios like
[SII]/H$\alpha$ and [NII]/H$\alpha$ \citep{mathis00, haffner09, blanc09}. Also clearly seen
in Figure \ref{fig-13} is a circum-nuclear star forming ring with a
diameter of $\sim 0.5$ kpc and a H$\alpha$ hole in the center of the
galaxy. These features and their physical origin will be discussed in
a future paper.

For each transition in Table \ref{tbl-6} we report the median $S/N$
over all spaxels in the data-cube and the fraction of the observed area
in which the emission line is detected at 5$\sigma$ and
3$\sigma$. Our data on NGC 628 reaches a 5$\sigma$ line flux limit per
spaxel (4 arcsec$^2$) of $\sim2\times10^{-17}$ erg s$^{-1}$ cm$^{-2}$.

As mentioned above, we detect H$\alpha$ at 5$\sigma$
over the full data-cube with a median $S/N=47$. 
Other transitions are usually detected at lower significance than
H$\alpha$. Individual members of the [NII]$\lambda \lambda$6548,6583 and
[SII]$\lambda \lambda$6717,6731 doublets, as well as H$\gamma$,
H$\beta$, [OIII]$\lambda$5007, and the blended [OII]$\lambda
\lambda$3726,3729 doublet are detected at 3$\sigma$ over more than $\sim$90\%
of the observed area, and the [OIII]$\lambda$4959
line is detected at 3$\sigma$ over 58\% of the area.

\subsection{Nebular Extinction from Balmer Decrement}

For the typical extinction regimes found in spiral
galaxies the H$\alpha$ to H$\beta$ flux ratio (a.k.a Balmer decrement) provides a
good estimate of the amount of dust extinction in the regions giving
rise to the nebular emission \citep[see discussion in][]{blanc09}. 
Assuming an intrinsic H$\alpha$/H$\beta$ ratio of 2.87 for Case B recombination
in $n=10^2$ cm$^{-3}$, $T=10^4$ K gas \citep{osterbrock06}, the observed ratio provides the amount of dust
reddening by means of the following relation

\begin{equation}
E(B-V) = \frac{-2.5\;{\rm log}\left( \frac{[{\rm H}\alpha/{\rm
        H}\beta]_{obs}}{2.87}\right)}{k(\lambda_{{\rm
      H}\alpha})-k(\lambda_{{\rm H}\beta})} 
\end{equation}

\noindent
where $[{\rm H}\alpha/{\rm H}\beta]_{obs}$ is the observed line ratio and
$k(\lambda)$ is the extinction law. We assume a foreground MW
extinction law as parameterized by \cite{pei92}. SMC and LMC laws were
tested \citep[also using the][parametrization]{pei92}, and no
significant change was observed in the derived extinction values
(these 3 extinction laws are practically identical at optical
wavelengths). 

\begin{figure}
\begin{center}
\epsscale{1.1}
\plotone{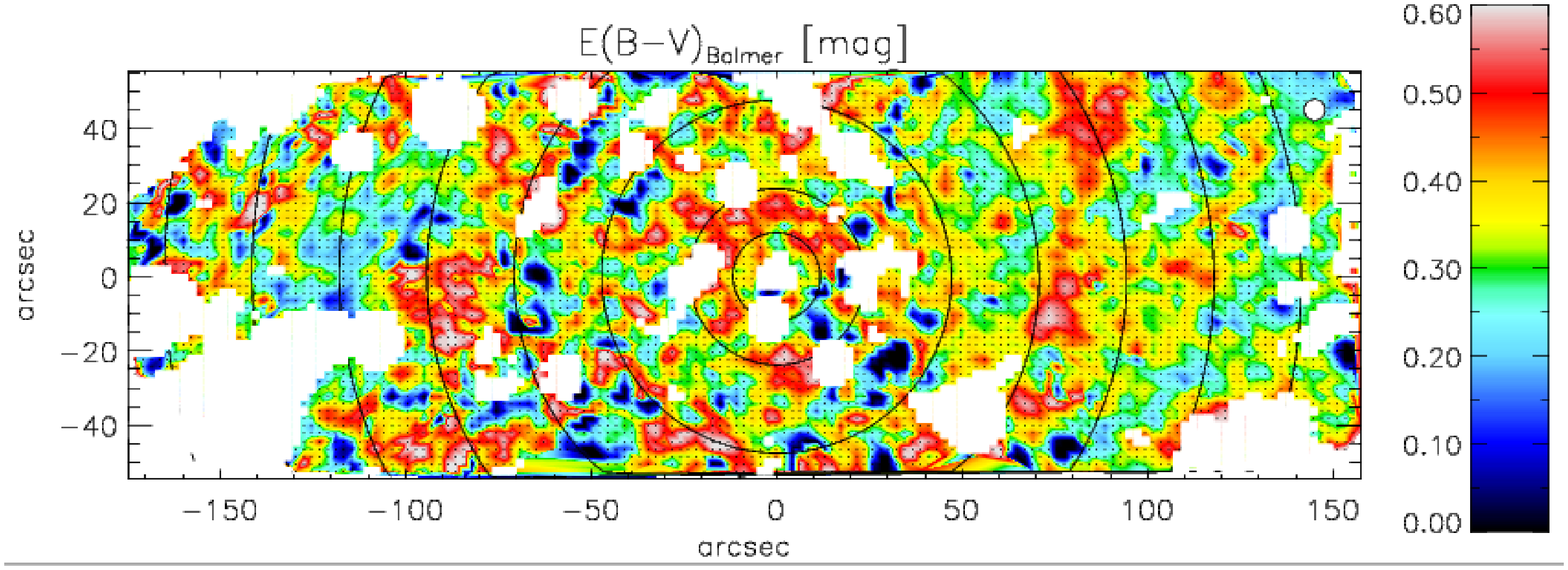}
\plotone{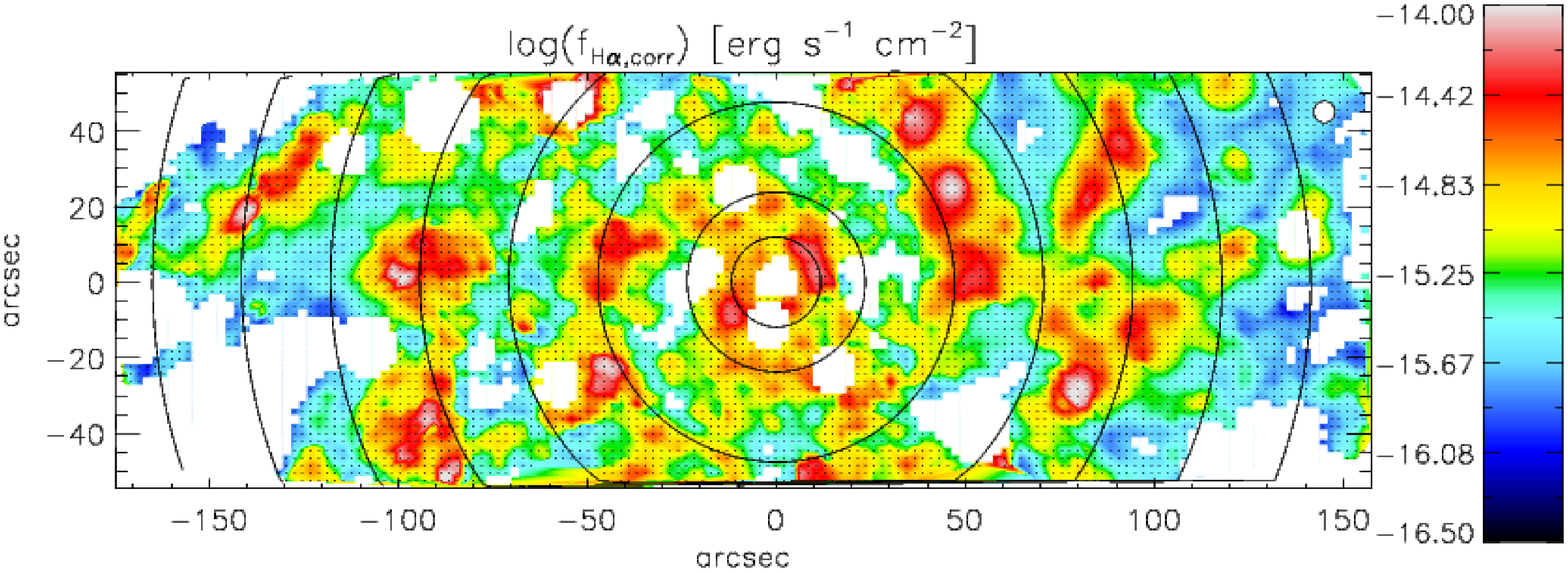}
\caption{{\it Upper Panel:} Map of the dust reddening $E(B-V)$ computed using
  Equation 4. Regions where the uncertainty in $E(B-V)$ is larger than
  0.2 mag have been masked. These masked regions correspond mainly to low surface brightness
  parts of the galaxy where the H$\alpha$ to H$\beta$ ratio cannot be measured
  with high $S/N$. {\it Lower Panel:} Map of the H$\alpha$ emission
  line flux corrected for the effects of dust extinction using a MW
  extinction law. Black contours are as in Figure \ref{fig-13}. Spaxels contaminated by
 foreground MW stars (see \S6.1) are masked in both panels.}
\label{fig-14}
\end{center}
\end{figure}

Figure \ref{fig-14} presents a map of $E(B-V)$
across the disk of NGC 628. Errors in the measured values of $E(B-V)$
are computed by propagating the uncertainty in the H$\alpha$ and
H$\beta$ line fluxes. In Figure \ref{fig-14} we have masked regions in
which the error in $E(B-V)$ is larger than 0.2 mag, which roughly amounts to
an uncertainty of $\sim50$\% in the extinction correction factor at the
wavelength of H$\alpha$. Regions of high extinction follow the inner
rims of the spiral structure (as traced by H$\alpha$), and a visual
comparison to CO maps of NGC 628 \citep{leroy08} shows almost perfect coincidence
between high dust extinction and high molecular gas surface density
regions. Also presented in Figure \ref{fig-14} is a map of
H$\alpha$ after correcting for dust extinction. The Balmer decrement
method implies a median extinction across all regions with good
$E(B-V)$ measurements of $A_V=1.06$ mag (assuming $R_V=A_V/E(B-V)=3.1$).

\begin{figure}
\begin{center}
\epsscale{1.2}
\plotone{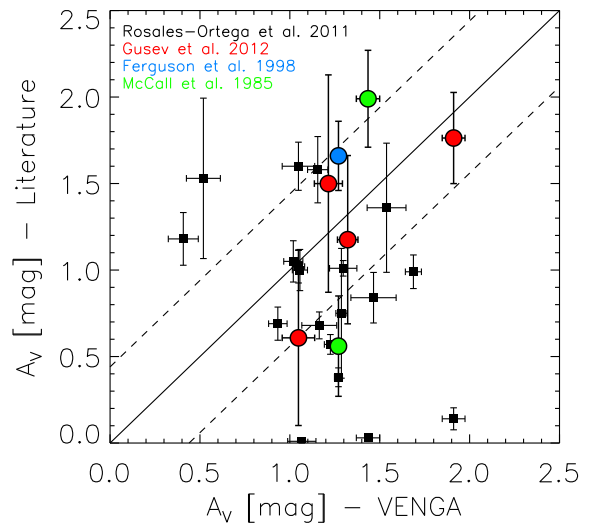}
\caption{Comparison between the dust extinction ($A_V$) derived from
  the Balmer decrement in VENGA and measruements found in the literature for a sample
  of HII regions in the \cite{mccall85, ferguson98, rosales-ortega11}
  and \cite{gusev12} samples. Dashed lines indicate a 50\% deviation in
the dust correction.}
\label{fig-15}
\end{center}
\end{figure}

In Figure \ref{fig-15} we compare the VENGA Balmer decrement dust extinction
measurements to a compilation of HII region measurements in the
literature. The VENGA datacube of NGC 628 covers 20 HII regions from the
catalog of \cite{rosales-ortega11} (built from the PINGS IFU datacube
of the galaxy), 4 HII regions from the multi-slit study of
\cite{gusev12}, and two HII regions from the \cite{mccall85}
sample, one of which was also studied by \cite{ferguson98}. The two
regions from the \cite{mccall85} sample were also detected by PINGS so
the total number of unique comparison HII regions in 24. In our
literature search we only considered regions with statistically
significant extinction measurements in the respective
studies.

Beacuse of the large spatial resolution of our data the comparison to
seeing limited slit spectroscopy measurements is challenging. Even the
PINGS data has a spatial resolution which is a factor
two smaller than ours. For the VENGA measurements we use the average
extinction in a $5.6''$ diameter aperture centered in each region,
expecting the nebular emission in that spatial resolution element to
be dominated by the HII region. Considering the differences in
methodology the comparison shows a reasonable agreement between the values measured in
VENGA and those measured by previous authors. The observed scatter is
consistent with the scatter seen in extinction among different
studies. For the one HII region which is common to the
\cite{rosales-ortega11}, \cite{mccall85}, and \cite{ferguson98}
studies the authors measure $A_V$ values of 0.38, 0.56, and 1.66 mag
respectively while the VENGA data yields an intermediate value of 1.27.

\section{Fitting the Stellar and Ionized Gas Velocity Fields}

In this section we present the methods used to fit the stellar and ionized
gas velocity fields of NGC 628. We are interested in
measuring the galaxy's rotation curve using both the stellar and
gaseous component with the goal of understanding any differences observed between
both velocity fields. We propose a new method to measure the
inclination of face-on disk galaxies which is based on matching the gas and
stellar rotation curves by means of the {\it asymmetric drift} correction.

\subsection{Removal of Foreground Stars}

\begin{figure}
\begin{center}
\epsscale{1.1}
\plotone{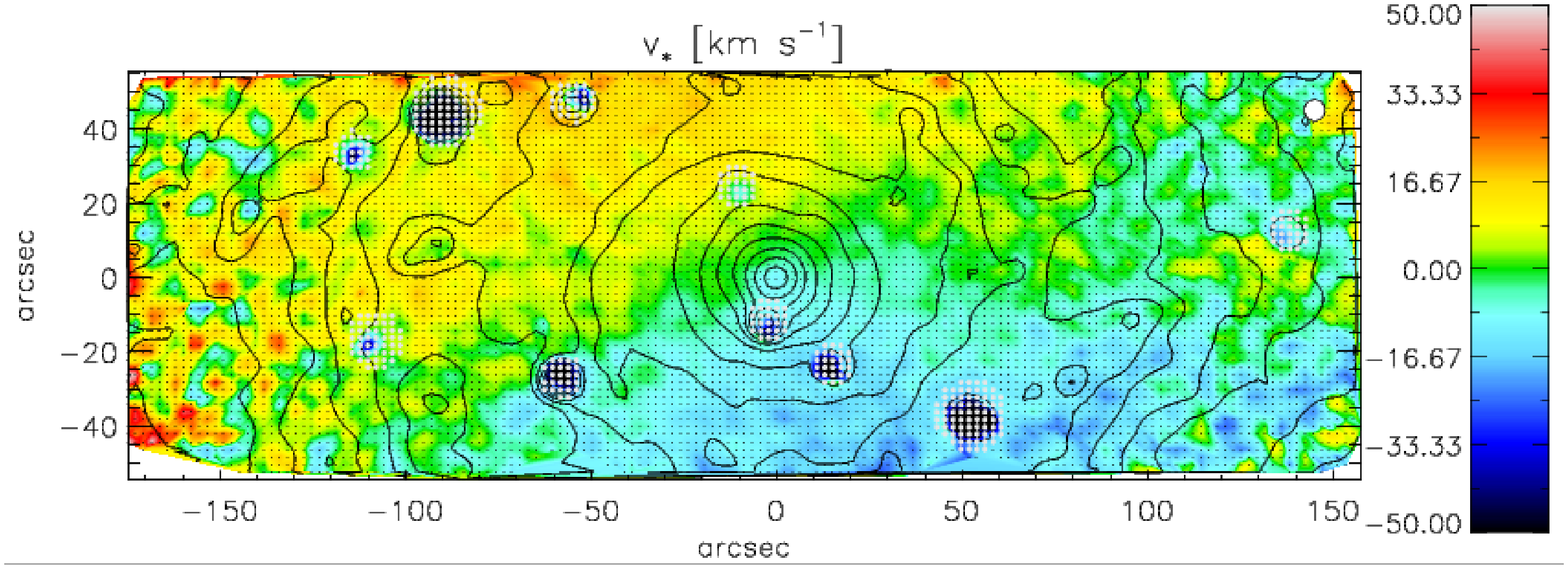}
\plotone{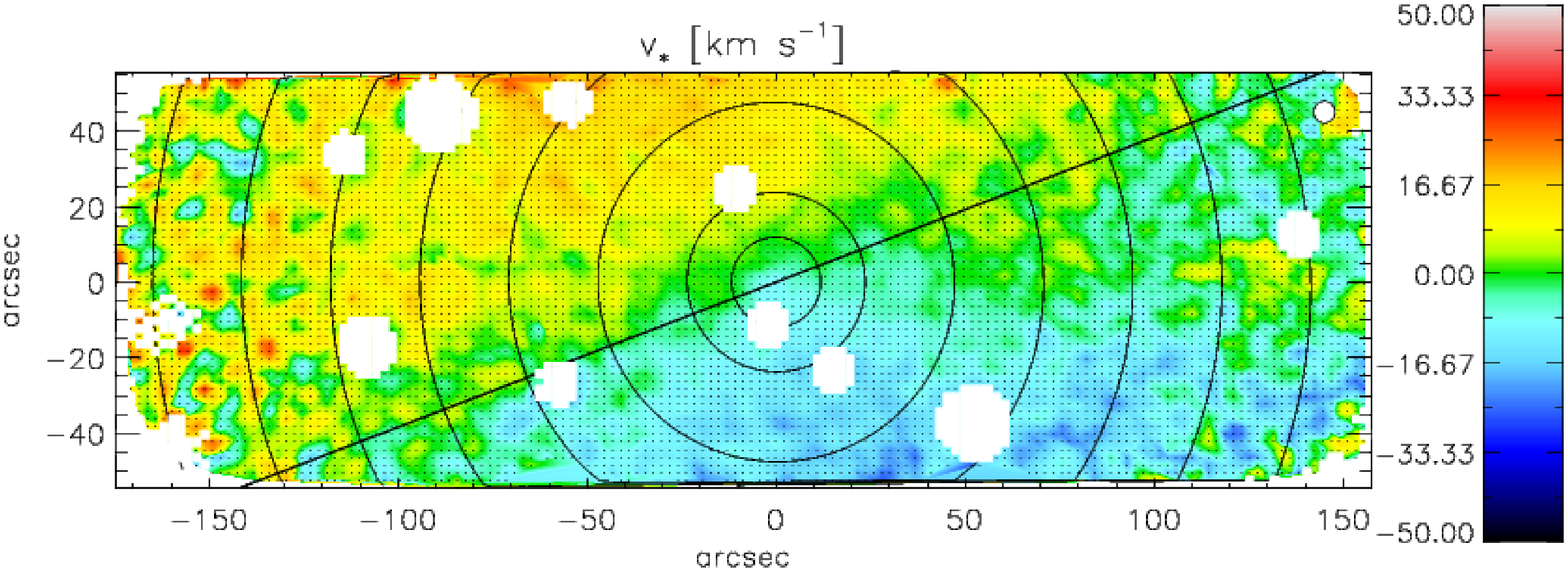}
\plotone{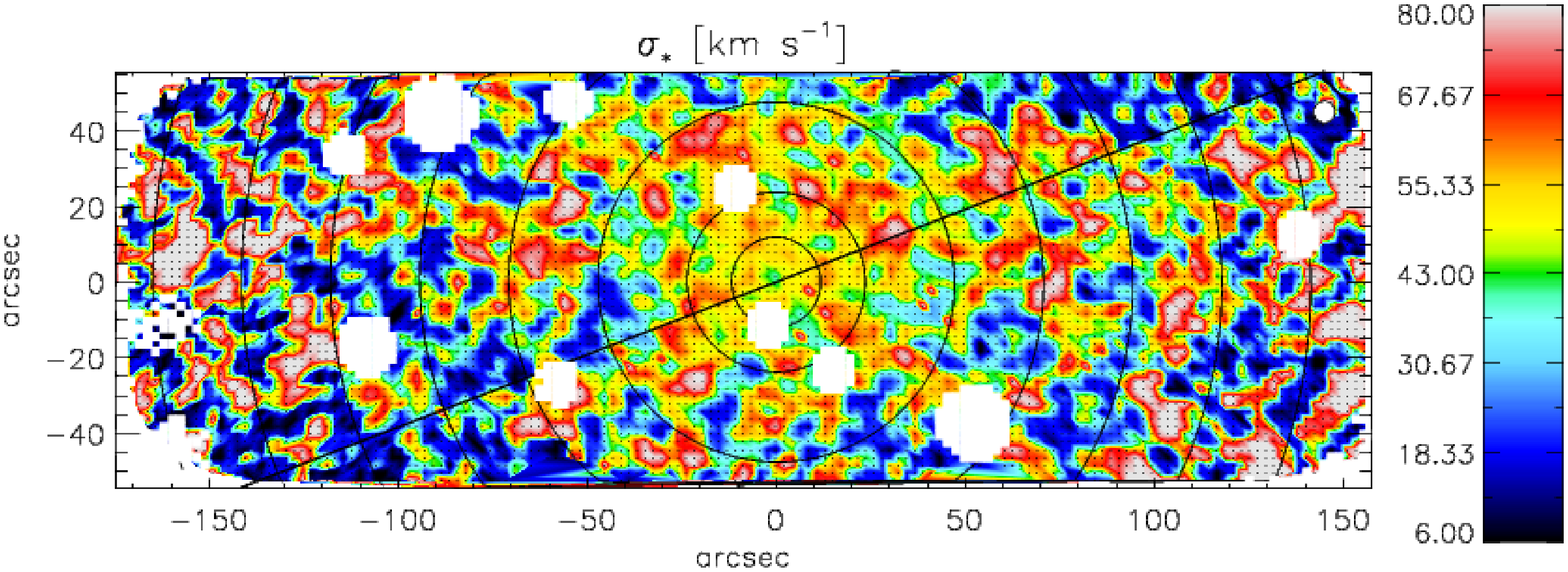}
\plotone{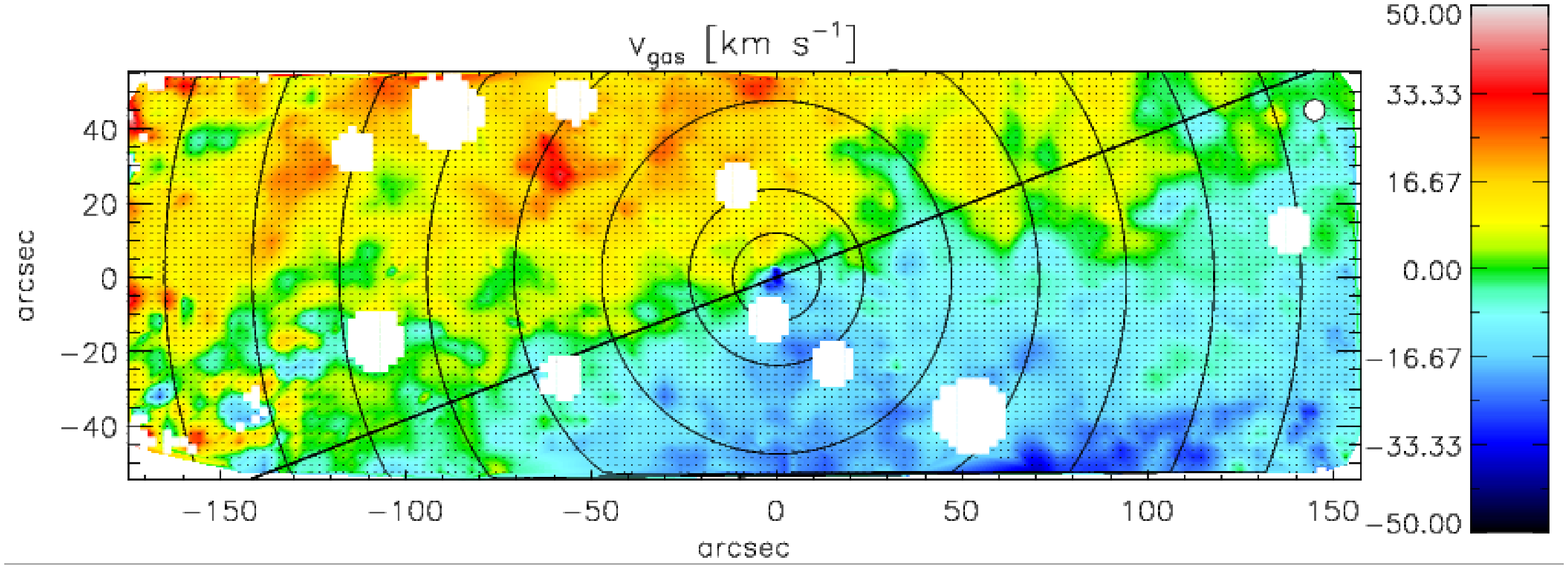}
\caption{{\it First Panel:} Map of the stellar LOS velocity ($v_{*}$) in NGC
  628. Black contours mark SDSS $r$-band isophotes as in Figure
  \ref{fig-9}. Gray dots mark spaxels flagged as contaminated by
  foreground stars. {\it Second Panel:} Clean map of the stellar LOS
  velocity where both foreground stars and regions with an error in
  $v_{*}$ larger than $15$ km s$^{-1}$ have been masked. {\it Third
    Panel:} Map of the LOS stellar velocity dispersion
  ($\sigma_{*}=\sigma_{\rm LOS}$) measured from the fit to the Mg I $b$
  region. {\it Lower
    Panel:} Map of the ionized gas LOS velocity ($v_{gas}$) as
  measured from the simultaneous fit to all emission lines listed in
  Table \ref{tbl-6}. Regions having an error in
  $v_{gas}$ larger than $15$ km s$^{-1}$ have been masked. The
  zero-velocity line for the measured 21$^{\circ}$ PA is marked by the
diagonal solid black line in the bottom three panels.}
\label{fig-16}
\end{center}
\end{figure}

In Figure \ref{fig-16} we present the stellar and ionized gas velocity
field of NGC 628 resulting from the fits described in \S5. Foreground
Milky Way stars can be clearly seen in the stellar velocity field as
sources of low velocity along the line of sight. This is evident by
looking at the SDSS $r$-band isophotes overploted on the top panel of
Figure \ref{fig-16}. Guided by the SDSS broad-band image, we flag these low
velocity spaxels in the data-cube as spectra which are significantly
contaminated by foreground stars. We mask all regions within $5''$ of these flagged pixels. 
The resulting stellar and ionized gas velocity fields after the masking of foreground
stars are shown in the second and bottom panels of Figure
\ref{fig-16}. Even at the low inclination of NGC 628 (see below for a
discussion on the actual value) the VENGA data allows us to clearly
detect the rotation of both the stellar and gaseous components of the
galaxy. 

\subsection{Kinematic Position Angle and Systemic Velocity}

We use the {\it symmetrization} method described in Appendix C of
\cite{krajnovic06} and implemented by Michele Cappellari in the IDL
routine FIT\_KINEMATIC\_PA\footnote{http://www-astro.physics.ox.ac.uk/$\sim$mxc/idl/}
to measure the kinematic position angle (PA) and systemic velocity ($v_{sys}$) from
both maps. In brief, for a given set of values (PA, $v_{sys}$), the
method assumes mirror symmetry in the velocity field and a
symmetrized velocity field is constructed by averaging the
absolute values of the velocity at mirroring points on opposite sides of the
perpendicular to the line of nodes. The best-fit PA and $v_{sys}$
are those which minimize the residuals between the observed and
symmetrized velocity fields.

For the kinematic PA we obtain consistent values of
$20\pm5^{\circ}$ and $22\pm6^{\circ}$ for the
stellar and ionized gas velocity fields respectively. This value is in
agreement with the RC3 $B$-band photometric PA of $25^{\circ}$
\citep{devaucouleurs91}, but is inconsistent with the 2MASS $K$-band
reported value of $87.5^{\circ}$ \citep{jarrett03}. The kinematic PA of
the ionized gas velocity field in the inner parts of the disk of
NGC 628 has been measured independently in the
past by \cite{daigle06} and \cite{fathi07} using Fabry-Perot
H$\alpha$ spectroscopy. These two studies measured inconsistent values
of $26.4\pm2.4^{\circ}$ and $15\pm5^{\circ}$. The value derived from
the VENGA data sits right in between these two measurements
and we consider it to be more reliable as it is based on deeper data,
using the kinematics of multiple emission lines, and shows good
agreement with the kinematic PA of the stellar component.

\cite{kamphuis92} used VLA HI 21 cm observations to reveal the
presence of an extended HI disk in NGC 628 reaching out to
a galactocentric radius of $\sim$30 kpc and showing complex
morphology and kinematic structure. The HI velocity field shows a
sharp transition in both PA and inclination around a galactocentric radius of $7'$
($\sim18$ kpc) believed to be caused by the presence of a warp in the outer HI
disk. The VENGA IFU data does not sample the disk out to such large galactocentric
radii but the stellar and ionized gas kinematics are consistent with
the HI kinematics within the inner disk of the galaxy. The HI velocity
field has a kinematic PA of $\sim25^{\circ}$ within the central $4'$
(Figure 6 in \cite{kamphuis92}), consistent with our measured value.

The symmetrization procedure also yields the systemic velocity of the
galaxy. For the gaseous component we obtain a value of $650\pm10$ km
s$^{-1}$, in agreement with the HI 21cm derived values of
$657\pm1$ km s$^{-1}$ \citep{kamphuis92} and $659\pm3$ km s$^{-1}$
\citep{walter08}, but in stark disagreement with the value of
$683\pm3$ km s$^{-1}$ derived by \cite{fathi07} from the H$\alpha$
velocity field. This inconsistency probably reflects a systematic error in the
parameters of the kinematic model in \cite{fathi07} who also report a
decreasing rotation curve within the central kiloparsec of the galaxy, 
feature which is not observed neither by \cite{daigle06} nor in this study.
The VENGA stellar velocity field yields a consistent systemic velocity of
$651\pm10$ km s$^{-1}$ also agreement with the HI
21cm derived values mentioned above.

\subsection{Harmonic Decomposition Modeling}

In order to measure the rotation curve of the galaxy we model the
observed stellar and ionized gas velocity fields using the harmonic
decomposition method \citep{franx94, wong04, krajnovic06}. In
particular we use the IDL KINEMETRY\footnote{http://www.eso.org/~dkrajnov/idl/\#kinemetry}
package to perform the fitting \citep{krajnovic06}. We fit the azimuthal dependance of the
LOS velocity along a set of elliptical annuli at different radii
along the semi-major axis of NGC 628. The annuli semi-major axes are
spaced following the combination of linear and logarithmic increase
factors described in \cite{krajnovic06} with a minimum radius of $6"$.
For each ellipse the LOS velocity is fit as

\begin{equation}
v(R,\psi)=v_{sys}+\sum\limits_{n=1}^3[A_n(R){\rm
  sin}(n\psi)+B_n(R){\rm cos}(n\psi)]  
\end{equation}

\noindent
where $v_{sys}$ is the systemic velocity measured
independently for each component in the previous section, $R$ is the
ellipse semi-major axis, and $\psi$ is the eccentric anomaly angle.
The outermost radius at which the fit is performed is set by the
requirement that data is available for 70\% of the points along
the ellipse. For NGC 628, this translates in an outermost radius of
$60''$ ($\simeq2.5$ kpc).

During the fitting, the geometric parameters of the ellipses (center, PA, and
inclination) are held fixed. We use the average of the two globally
derived kinematic PAs for stars and gas of
$21^{\circ}$ (\S6.2), and the central coordinates from NED\footnote{The
  NASA/IPAC Extragalactic Database (NED) is operated by the Jet
  Propulsion Laboratory, California Institute of Technology, under
  contract with the National Aeronautics and Space
  Administration.}. As discussed below, the choice of an
inclination value for NGC 628 is far from trivial.

Using photometric axis ratios ($q=b/a$) to compute the inclination of NGC 628
by adopting the \cite{tully88} method we derive $i=28^{\circ}$ ($q=0.91$)
for the RC3 \mbox{$B$-band} measurements \citep{devaucouleurs91} and
$i=34^{\circ}$ ($q=0.86$) for the 2MASS $K$-band photometry \citep{jarrett03}.
The detailed photometric decomposition of \cite{fisher08} implies an
inclination of $i=33^{\circ}$ ($q=0.87$) for the disk component. 
Given the projected circular velocities measured at large radii in
both the 21cm HI and H$\alpha$ velocity fields \citep{kamphuis92, daigle06, fathi07},
such inclinations would imply a deprojected velocity of \mbox{$v_{flat}\sim50$ km s$^{-1}$} for
the flat part of the rotation curve. Such a small circular velocity
is wildly inconsistent with expectations from the Tully-Fisher (TF)
relation \citep{tully77} for NGC 628. 

Using the $K$-band TF relation from \cite{verheijen01}\footnote{We use
  the parameters from the fit to their ``RC/FD Sample, Excluding NGC
  3992'' in Table 4 of \cite{verheijen01}} and the 2MASS $K$-band
absolute magnitude from Table \ref{tbl-1} we obtain
$v_{flat,TF}=149\pm3\pm8$ km s$^{-1}$, where the random error comes
from the uncertainty in the distance and apparent magnitude of the
galaxy and the systematic uncertainty comes from the 0.26 mag scatter
in the adopted TF relation. The inconsistency between this value and
that derived from the photometrically measured inclination is not
surprising given that the uncertainty in $i$ explodes as one approaches
a nearly face-on configuration. This is also the case for kinematically
derived inclinations. In fact, letting the inclination be a free
parameter for the ellipses in our harmonic decomposition fits yields an unrealistic
median value of $i=45^{\circ}$. \cite{daigle06} derives a
kinematic inclination of $21.5\pm4.5^{\circ}$ from their tilted ring
analysis of the H$\alpha$ velocity field. For an excellent discussion on the
uncertainties associated with measuring the inclination of disk galaxies
see \cite{bershady10b}.

Given the unreliability of both photometric and kinematic
inclinations for face-on galaxies, most studies of NGC 628 have assumed a value of
$i=6.5^{\circ}$ \citep[e.g. ][]{kamphuis92, daigle06, fathi07,
  herbert-fort10}. This value was adopted by
\cite{kamphuis92} so the flat part of the HI rotation curve would
reach an arbitrary chosen value of $v_{flat}=200$ km s$^{-1}$. A
notable exception is the work of \cite{herrmann08} who fit an
exponential model to the THINGS 21 cm HI velocity field of NGC 628 to obtain an
inclination of $i=9^{\circ}$.

Under the assumption that NGC 628 exactly follows the TF relation, a better value for
the inclination would be $i_{TF}=8.7\pm0.7^{\circ}$ (assuming an error of
10 km s$^{-1}$ for the value of $v_{flat}$ measured at large radii by
\cite{kamphuis92} and propagating the errors in the apparent
magnitude and distance of NGC 628 as well as the scatter in the TF
relation). This so called ``Tully-Fisher inclination'' method is
discussed in \cite{bershady10b} and provides more reliable estimates of
the inclination for face-on systems than photometric or kinematic
techniques do.

In the following section we propose a new alternative method to derive the
inclination of face-on disk galaxies by demanding agreement between
the ionized gas and stellar rotation curves after applying the
asymmetric drift correction to the stellar component.

\subsection{Rotation Curve and Asymmetric Drift Correction: A New
  Method to Measure Inclinations}

Because of the collisional nature of gas, as opposed to the
collisionless nature of stars, the cold ISM of disk galaxies is
expected to be confined to a relatively thin disk supported mainly by
rotation. Although pressure support in the
ISM can become important under certain circumstances (e.g. high
turbulence, strong pressure gradients, etc.), the small and relatively constant
H$\alpha$ velocity dispersion of $\sigma_{H\alpha}\sim17$ km s$^{-1}$
measured by \cite{fathi07} across the disk of NGC 628 implies that the
gaseous disk of this galaxy is almost fully supported by circular
motions, with random motions only having a minor effect. The importance of
pressure support against gravity in gaseous galactic disks is
extensively discussed in \cite{dalcanton10}. Using the flat dispersion
profile of \cite{fathi07} and applying equation 11 from
\cite{dalcanton10} we compute a correction to
the ionized gas rotation curve of NGC 628 of $\sim
2$ km s$^{1}$. This is many times smaller than the 1$\sigma$ uncertainty in
the rotation curve itself. Given the small magnitude of this effect we decide not
to apply these corrections and neglect the effect of pressure support
in the gaseous disk of NGC 628. Therefore we consider the
ionized gas rotation curve to roughly follow the dynamics of a test particle under
the gravitational potential in the mid-plane of the galaxy.

Stellar disks can also experience support against
gravity provided by random motions. As in the case of pressure
support in gaseous disks, stellar random motion support translates in
a measured circular velocity which is smaller than the actual circular
velocity at which a test particle would orbit in the gravitational
potential of the galaxy. This effect, first noted as a difference in
orbital velocity for stellar populations of different ages (and
therefore different velocity dispersions) in the solar neighborhood
is known as ``asymmetric drift'' \citep{stromberg22}. As noted by
\cite{dalcanton10} the equations describing it are similar to those
describing pressure support in gaseous disks but the physical
phenomena are fundamentally different. 

Starting from the Jeans equations \citep{jeans15} and following
\cite{binney87}, for a stellar axisymmetric disk in steady
state, under the assumption that the ``epicyclic approximation''
holds and that the stellar velocity ellipsoid (SVE) is aligned with
the disk in cylindrical coordinates and its shape is independent of
the vertical distance to the plane, the quadratic difference between
the true and the observed circular velocities is given by

\begin{equation}
 v_c^2-v_{\phi}^2=\frac{\sigma_R^2}{2} \left (\frac{\partial
    {\rm ln}(v_{\phi})}{\partial {\rm ln}(R)} -1 -2\frac{\partial {\rm ln}(\nu
    \sigma_R^2)}{\partial {\rm ln}(R)}\right )
\end{equation}

\noindent
where $v_c$ is the circular velocity of a test particle in
the potential, $v_{\phi}$ is the observed circular velocity,
$\sigma_R$ is the velocity dispersion in the radial direction, and $\nu$
is the number density of stars. While $v_{\phi}$ and its logarithmic
derivative with radius can be directly measured from the harmonic
decomposition fits described in \S6.3, calculating the last term in
the perenthesis in Equation 6 requires making a series of assumptions
which we describe below.

In order to apply the asymmetric drift correction we
assume a constant scale height and $V$-band mass-to-light ratio for the
disk of NGC 628. These assumptions imply that $\nu(R)$ is proportional
to the stellar mass surface density $\Sigma(R)$ and that the latter is well
traced by the surface brightness profile of the galaxy. In particular we use the best-fit
disk plus pseudo-bulge $V$-band decomposition of \cite{fisher08} to
trace the radial dependance of $\nu(R)$ when calculating the last term
in Equation 6. Note that knowledge of the $V$-band mass-to-light ratio
is not necessary as it cancels out in the logarithmic derivative.

\begin{figure}
\begin{center}
\epsscale{1.2}
\plotone{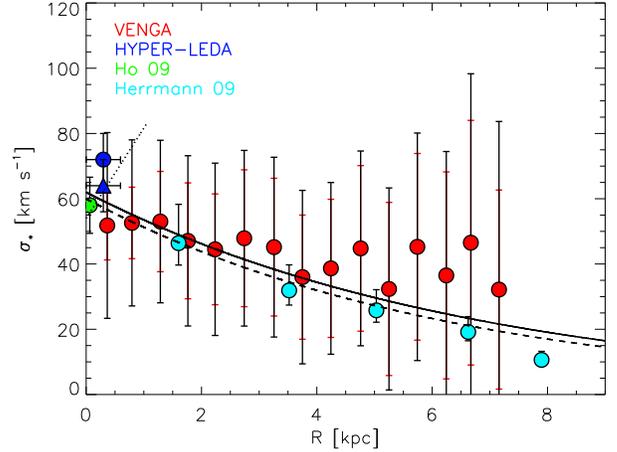}
\caption{Radial profile of 
$\sigma_{*}=\sigma_{LOS}$. Red circles mark the
median $\sigma_*$ for all spaxels in 0.5 kpc wide radial bins. Error bars show both
the standard deviation in each radial bin (red) and the median error
in $\sigma_*$ for each spaxel (black). The central stellar velocity dispersion values
from HyperLeda and \cite{ho09} are shown in blue and green respectively.
The dispersion profile measured by \cite{ganda06} using the SAURON IFU
spectroscopy is shown as a black dashed curve. The velocity dispersion
profile measured from PN kinematics by \cite{herrmann08} is also
shown (light blue circles, dashed line). The solid black curve
shows the best-fit dispersion profile to the VENGA data used to calculate the asymmetric drift
correction.}
\label{fig-17}
\end{center}
\end{figure}

Using a measurement of the stellar velocity dispersion along the line of sight
($\sigma_{\rm LOS}(R)$) we can estimate $\sigma_R(R)$, provided our
assumptions regarding the shape and orientation of
the SVE. The third panel in Figure \ref{fig-16} presents the map of
the stellar velocity dispersion ($\sigma_*=\sigma_{\rm LOS}$) measured from the VENGA data in
\S5.1. In the central regions of NGC 628 we see $\sigma_*$ falling
from a central value of $\sim60$ km s$^{-1}$. Figure \ref{fig-17} presents the radial
profile of $\sigma_{\rm LOS}$. To measure the radial profile we only
considered spaxels with high enough $S/N$ as to
provide a reliable fit of the LOSVD (i.e. those which are not masked
in Figure \ref{fig-16}). Red circles mark the median $\sigma_{\rm LOS}$ in 0.5
kpc wide radial bins. Error bars show both the standard deviation in
each radial bin (red) and the median error in $\sigma_*$ for each
spaxel (black).

Also plotted in Figure \ref{fig-17} is the central stellar velocity
dispersion value of $\sigma_{*}(R=0)=72.2\pm7.8$ km s$^{-1}$ from
HyperLeda\footnote{http://leda.univ-lyon1.fr/} (blue circle). The
HYPERLEDA value is an average over six literature measurements and is
driven by an outlier at 116 km s$^{-1}$ \citep{dinella95}. Rejecting
this outlier the HYPERLEDA average comes down to $64$ km s$^{-1}$, in
even better agreement with our measurements (blue triangle in Figure
\ref{fig-17}). The velocity dispersions we measure in the central regions
of the galaxy are also consistent with the nuclear value of
$58.0\pm8.6$ km s$^{-1}$ measured by \cite{ho09} (green circle). Our
measurements also agree with the central dispersion of $54\pm10$
km s$^{-1}$ measured by \cite{ganda06} using SAURON IFU
spectroscopy. The radial $\sigma_{*}$ profile measured in this last
study is shown as a dashed black line in Figure \ref{fig-17}. While
our measured values in the center of the galaxy are consistent with
the ones in \cite{ganda06} we find no evidence to support their claim
of an increasing velocity dispersion with radius in NGC 628.
At larger radii Figure \ref{fig-17} shows the dispersion profile measured
by \cite{herrmann08} using the kinematics of a large sample of planetary
nebulae (PN). The dashed line shows the best fit dispersion profile
from \cite{herrmann08} which follows an exponential with a scalength
aproximately equal to two times the photometric scalelength of the
galaxy. The latter agrees with theoretical expectations for constant
scaleheight disks in which $\sigma_z(R) \propto
\sqrt{\Sigma_{disk}(R)}$, with the proportionality constant depending
on the detailed vertical profile of the disk \citep[exponential, sech or
sech$^2$, ][and references therein]{vanderkruit88, bershady10, vanderkruit11}.

The 145 km s$^{-1}$ spectral resolution at Mg I $b$ in
the VENGA VIRUS-P data makes the measurement such low values of the
velocity dispersions very challenging. The large error bars in Figure
\ref{fig-17} indicate that for single spectra individual measurements of low dispersion values ($\sim 50$ km
s$^{-1}$) are extremely noisy even at the high $S/N\sim100$
of our data. Nevertheless, by averaging over many
spaxels in the data-cube $\sigma_{LOS}$ we seem to recover the
velocity dispersion profile of the galaxy at least out to $R\simeq4$
kpc where the VENGA measurements flatten around a value of $\sim40$ km
s$^{-1}$. Comparison to the dispersion profile derived from PN
kinematics shows that this flattening is not real and most likely
corresponds to a systematic floor caused by uncertainties associated
with template missmatch and deviations from the assumed Gaussianity in
the instrumental line spread function most likely set this limit. This
is not surprising given that it happens at about a fourth of the
instrumental resolution and that a similar behavior is seen in the
simulations presented in \S5.2.

We fit an exponential dispersion profile to the VENGA data in the
$1.5<R<4$ kpc region in order to avoid regions in which the stellar
dynamics are contaminated by the central spheroidal component
\citep[see discussion in ][]{bershady10b}. We fix the scale length of
the profile to two times the scale length of the disk component in
the \cite{fisher08} decomposition. The
best fit profile is shown as a solid line in Figure \ref{fig-17},
implies a central velocity dispersion of $\sigma_{\rm LOS}(0)=62$ km
s$^{-1}$ (consistent with the HYPERLEDA, \cite{ho09} and
\cite{ganda06} values) and is in excellent agreement with the
\cite{herrmann08} best fit profile and with theoretical expectations
for the dynamics of stellar disks. This $\sigma_{\rm LOS}(R)$ profile
is used to derive $\sigma_{\rm R}(R)$ in order to calculate the
assymetric drift correction using Equation 6.

Following the method outlined in \cite{bershady10b} we assume the
following shape for the SVE in order to deproject the LOS velocity
dispersion profile and compute $\sigma_R(R)$:

\begin{equation}
\alpha=\frac{\sigma_z}{\sigma_R}=0.6
\end{equation}
\begin{equation}
\beta=\frac{\sigma_{\phi}}{\sigma_{R}}=\sqrt{\frac{1}{2}(\frac{\partial{\rm ln}(v_{\phi})}{\partial {\rm ln}(R)}+1)}
\end{equation}

where the first equation comes from expectations from the
solar neighborhood \citep{binney98} and observations of nearby
galaxies \citep{bershady10b} and the second comes from adopting the
epicyclic approximation. This allows us to compute all the terms in
Equation 6 and apply the asymmetric drift correction to the stellar
velocity field of the galaxy. We use Monte Carlo simulations to
calculate the error in the asymetric drift correction. In the
error calculation we include the uncertainty in the observed rotation
curve which is propagated to compute the uncertainty in the
logarithmic radial derivative of the rotation curve and the value of $\beta$, an error of
$\Delta\alpha=0.15$ \citep{bershady10b}, and an error in $\sigma_{\rm
  LOS}(0)$ of 10 km s$^{-1}$. 

\begin{figure}
\begin{center}
\epsscale{1.1}
\plotone{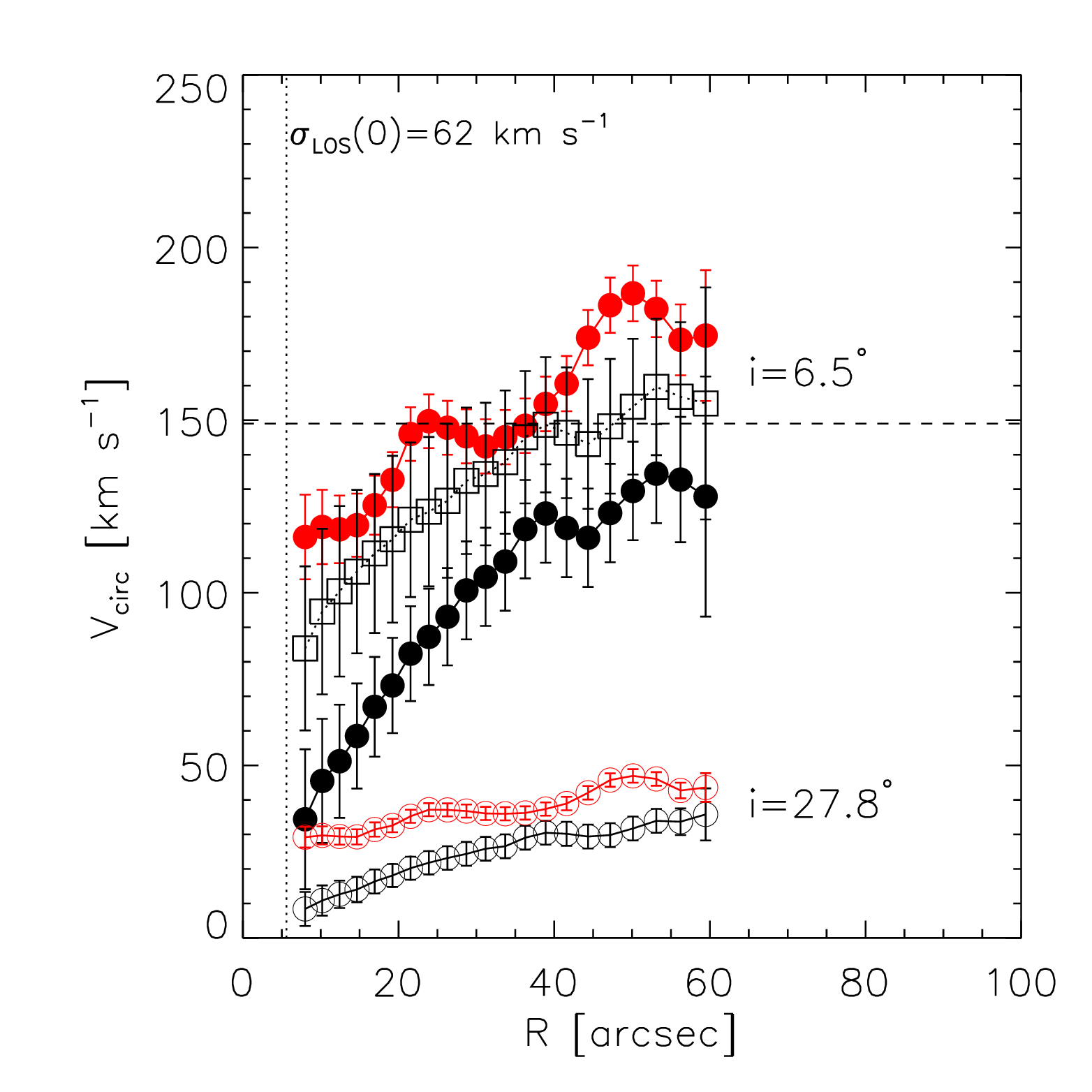}
\plotone{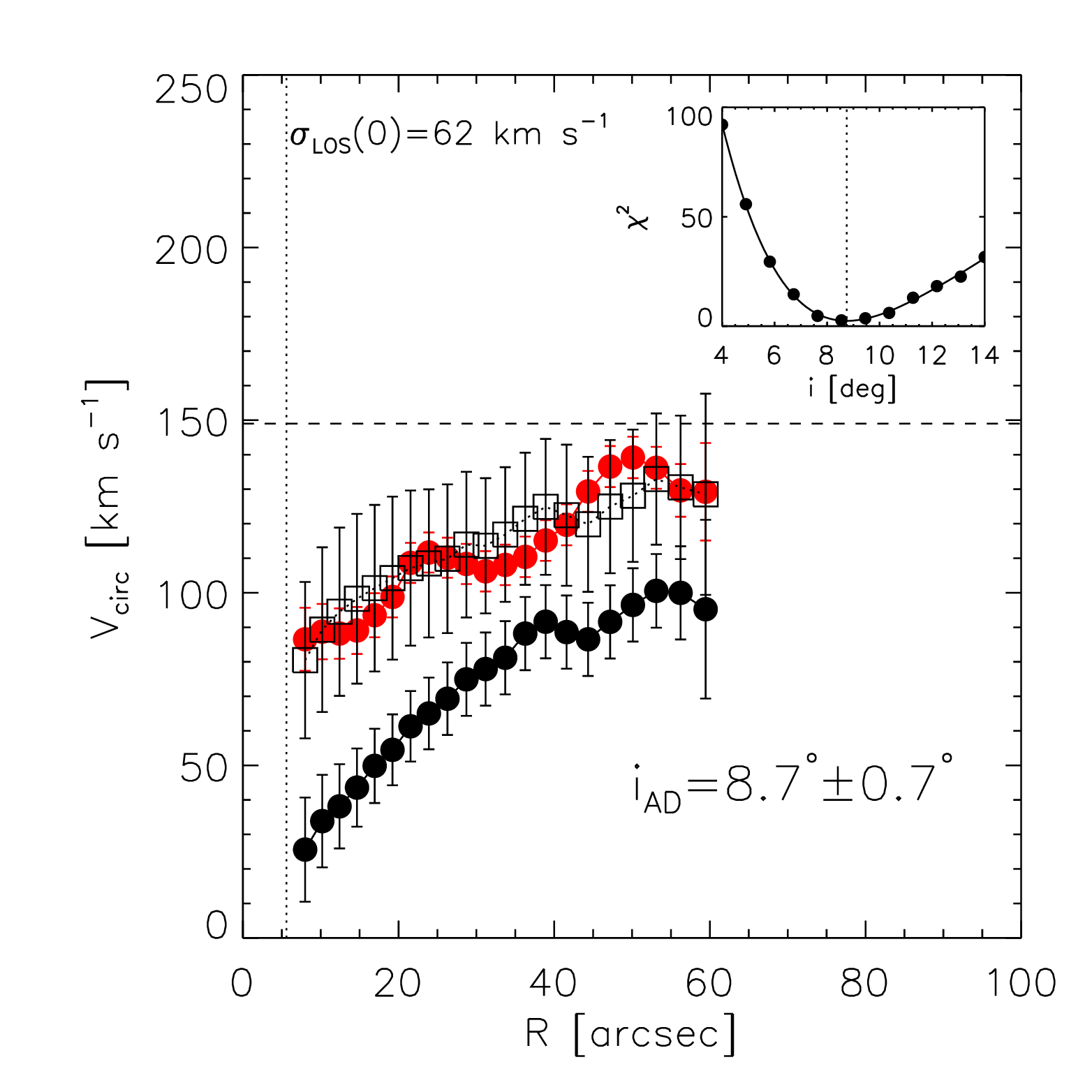}
\caption{{\it Top Panel:} Stellar (black) and ionized gas (red)
  rotation curves for NGC 628 under the assumption of two different
  inclination values of 6.5$^{\circ}$ (filled circles) and
  27.8$^{\circ}$ (open circles). Open squares show the stellar
  rotation curve after applying the asymmetric drift correction for
  the $i=6.5^{\circ}$ case. The horizontal dashed line marks the
  asymptotic rotation velocity of $149$ km s$^{-1}$ expected from the
  Tully-Fisher relation. The size of the VENGA PSF FWHM is marked by the
  dotted vertical line.  {\it Lower Panel: } Same as above, but for
  the best-fit {\it asymmetric drift inclination}
  ($i_{AD}=8.7^{\circ}\pm0.7$). The inset shows the $\chi^2$
  distribution for the range of inclination values explored during the
  fit. The vertical dotted line in the inset marks the best-fit inclination.}
\label{fig-18}
\end{center}
\end{figure}

In the upper panel of Figure \ref{fig-18} we present the VENGA
rotation curve of NGC 628 for both stars (black circles) and ionized
gas (red circles), as measured
from the first order cosine term of the harmonic decomposition
model $v_{circ}(R)=B_{1}(R)/sin(i)$. We show measurements of the rotation curve
for two different assumed inclination values of $i=28^{\circ}$ (derived from the RC3 axis
ratios, open circles) and $i=6.5^{\circ}$ \citep[][, filled
circles]{kamphuis92}. The horizontal dashed line marks the expectation
from the TF relation $v_{flat,TF}=149$ km s$^{-1}$ (\S6.3). The
high inclination implied by the photometric axis ratio dramatically
underestimates the amount of rotation expected in NGC 628. This
highlights the lack of reliability of geometrically measured
inclinations in face-on systems. 

Independent of the assumed inclination, the observed stellar
circular velocity (before applying the asymmetric drift correction) is
systematically lower than the observed circular velocity of the
gaseous component. This implies the
presence of significant kinematic support arising from random motions
in the stellar disk of NGC 628. For the $i=6.5^{\circ}$ case
the corrected stellar rotation curve after accounting for asymmetric
drift using equation 6 is shown by the black open squares in the top
panel of Figure \ref{fig-18}. Errorbars in the corrected stellar
rotation curve are calculated using Monte Carlo simulations and
including the uncertainty in all the quantities entering the correction.
For an assumed value of $i=6.5^{\circ}$ the magnitude of the asymmetric drift correction
is not sufficient to account for the discrepancy between the observed
stellar and gaseous rotation curves. 

The above problem can be solved by changing the assumed value for the
the inclination. We remind the reader that the $i=6.5^{\circ}$ value
typically used in the literature for NGC 628 was arbitrarily adopted
by \citep{kamphuis92} to make the flat part of the rotation curve
reach a velocity of 200 km s$^{-1}$. From the TF relation we have
seen in \S6.3 that the expected asymptotic velocity in the flat part
of the rotation curve of NGC 628 is smaller than what was assumed by
\cite{kamphuis92} with a value closer to $150$ km s$^{-1}$. This
asymptotic velocity implies an inclination of  $i_{TF}=8.7^{\circ}$.

The lower panel in Figure \ref{fig-18}, shows the result of matching
the gas and asymmetric drift corrected stellar rotation curves to
obtain the inclination $i_{\rm AD}$ of the galaxy. The method consists
of sampling a range of inclination values in order to find the
inclination which minimizes the residuals between both rotation
curves. For each inclination value we measure the rotation curves by
performing harmonic decomposition fits to the velocity fields, we
deproject the LOS velocity dispersion profile to compute the
asymmetric drift correction, and we compute a $\chi^2$ statistic in
order to quantify deviations between the gas and corrected stellar
rotation curves.

The inset in the lower panel of Figure \ref{fig-18} shows the $\chi^2$
distribution as a function of inclination. The minimum $\chi^2$
corresponds to a best-fit value for the {\it asymetric drift inclination}
of $i_{\rm AD}=8.7\pm0.7^{\circ}$ in perfect agreement with the $i_{TF}=8.7\pm0.7^{\circ}$ value derived
from the TF relation. The exact agreement in both the value and the
error-bars implies that both methods produce consistent results with
similar precisions, at least in the case of NGC 628. The TF and
asymmetric drift methoda are highly complementary. The TF method
requires knowledge of the observed rotation velocity in the flat part
of the rotation curve at large radii, while the asymmetric drift method can be used
even when only kinematic measurements are only available in the
central regions of galaxies. Furthermore, the measurement of an independent {\it
  asymetric dirft inclinations} can allow for highly inclined systems
to be included when measuring the TF relation. 

We propose the matching of the gaseous and asymmetric drift corrected
stellar rotation curves as a new method to estimate the inclination of
nearly face-on disk galaxies. The fact that a very small difference of
2.2$^{\circ}$ in the adopted inclination can produce a catastrophic
inconsistency between the observed gas and stellar kinematics as in
the $i=6.5^{\circ}$ case highlights the sensitivity of this method as tool to measure
inclinations.

\section{Summary and Conclusions}

In this work we have presented the survey design, sample,
observing strategy, data reduction, and spectral analysis methods for
VENGA. Wide field integral field spectroscopy
proves to be a powerful tool to study a large set of physical phenomena
occurring in nearby galaxies. We characterized our sample of 30 nearby
disk galaxies in terms of their stellar masses and $SFR$s, and showed
that VENGA is a representative sample of massive ($>10^{10}$
M$_{\odot}$) spiral galaxies in the local universe. A large range of
morphological parameters (bulge-to-disk ratio, bar presence, bulge
S\'ersic index, inclination) is represented in the sample. The distance
distribution of our targets implies a median spatial resolution in
physical units of 383 pc. Our targets typically span a few arcminutes
in angular size, and we tile the VIRUS-P IFU in order to provide
a coverage out to 0.7 $R_{25}$ for most galaxies. This implies that
the VENGA data-cubes are among the largest optical IFU data-cubes ever
constructed. In particular, the NGC 628 data-cube presented in this
paper spans $5.2'\times1.7'$.

The wealth of information produced by
integral field spectrographs stresses the need for optimized
and pipelined software tools to process and analyze the
data. This becomes essential when this observational technique is
used to conduct large surveys like the one presented here.
We have presented the reduction and calibration pipeline used
for the VENGA data. We have also described our spectral analysis
pipeline which we use to extract stellar and gas kinematics as well
as emission line fluxes, and we have discussed the methods used to
estimate the uncertainties in all these parameters. When possible, we
have adapted existing publicly available software to be used on the VIRUS-P data.
We assessed the quality of the VENGA data obtained on NGC 628 and we find
it to be excellent. VIRUS-P provides high S/N spectra for
single fibers out to large galactocentric radii. Thanks to the ability
of reconstructing broad-band images from the IFU data and
cross-correlating them with archival broad-band images of the galaxies
in our sample we can achieve good astrometric precision ($\sim0.1''$)
and better than 10\% flux calibration accuracy across the full
wavelength range.

Finally we have used an harmonic decomposition technique to fit the
stellar and ionized gas velocity fields in NGC 628. The observed
kinematics imply that the gaseous disk of the galaxy (as traced by
nebular emission from ionized gas) is almost fully supported by
rotation while the stellar disk shows significant levels of
dynamic support arising from random motions. This is evident by a
$\sim50$ km s$^{-1}$  offset between the ionized gas and stellar
rotation curves.

We have shown that the above discrepancy in the rotation curves can be
explained if one takes into account the effect of stellar random
motions by applying an asymmetric drift
correction. The agreement between the ionized gas rotation curve and
the asymmetric drift corrected stellar rotation curve depends strongly
on the assumed value for the inclination. We take advantage of this
dependance and propose this as a method
to estimate the actual inclination of nearly face-on disk galaxies
for which classically used geometrically derived inclinations are
extremely unreliable. Applying this method we find a revised
asymmetric drift inclination for NGC 628 of $i_{AD}=8.7\pm
0.7^{\circ}$, in excellent agreement with the expected value of
$i=8.7\pm0.7^{\circ}$, which comes from assuming that NGC 628
satisfies the local TF relation.  

The main purpose of this work is to present the VENGA survey and
some preliminary scientific results have been discussed only
briefly. We expect to use the VENGA sample to conduct a series of
studies on star formation, galactic structure
assembly, stellar populations, chemical evolution, galactic feedback,
nuclear activity, and the properties of the interstellar medium in massive
disk galaxies, and to present these results in a series of future publications.
As an example of the practical uses of this data, in a
recent paper we have presented a study of the radial profile in the CO
to H$_2$ conversion factor $X_{\rm CO}$ and its relation to other parameters like the metallicity, the gas
surface density and the ionization parameter across the disk of NGC 628
\citep{blanc12}. 

Once all the VENGA data is processed we expect to make it publicly
available to the community.  The richness of large IFU datasets like the one we are compiling goes beyond the
scientific goals of our team. We expect VENGA to be a useful
resource that will complement the wealth of multi-wavelength datasets
that the community has acquired on nearby spiral galaxies over the last few
decades.

The VENGA collaboration acknowledges the generous support from the
Norman Hackerman Advanced Research Program (NHARP)
ARP-03658-0234-2009, GAB. acknowledges the support of Sigma Xi,
The Scientific Research Society, Grant in Aid of
Research. NJE and ALH were supported by NSF Grant AST-1109116. RXL and
LH acknowledge the supports by the National
Natural Science Foundation of China under grant No. 11073040, by 973 Program of China under
grant No. 2009CB824800, and by Shanghai Pujiang Talents Program
under grant No. 10pj1411800. The construction of the Mitchell Spectrograph (formerly VIRUS-P) was 
possible thanks to the generous support of the Cynthia \& George
Mitchell Foundation. We thank Phillip McQueen and Gary Hill for designing and constructing
VIRUS-P, and for their advice on the use of the instrument. We also
acknowledge David Doss and the staff at McDonald Observatory for their
invaluable help during the observations. This research has made use of NASA's Astrophysics
Data System, and of the NASA/IPAC Extragalactic Database (NED) which
is operated by the Jet Propulsion Laboratory, California Institute of
Technology, under contract with the National Aeronautics and Space
Administration.

\clearpage
\begin{deluxetable*}{llcccccccccc}

\tabletypesize{\tiny}
\tablecaption{The VENGA Sample\label{tbl-1}}
\tablewidth{0pt}
\tablehead{
\colhead{Object} & 
\colhead{Equatorial Coord.\tablenotemark{a}} & 
\colhead{Type\tablenotemark{a}} & 
\colhead{$i$\tablenotemark{a}} & 
\colhead{$\theta$\tablenotemark{a}} &
\colhead{$d_{25}$\tablenotemark{a}} &
\colhead{$D$} &
\colhead{Method\tablenotemark{d} } &
\colhead{pc/''} &
\colhead{$M_K$\tablenotemark{e}} &
\colhead{$\mu_B$\tablenotemark{a}} &
\colhead{N$_{P}$}
}

\startdata
 & $\;\;\;\;\;\;\alpha\;\;\;\;\;\;\;\;\;\;\;\;\;\;\;\delta$ & 
 &
deg &
deg &
arcmin &
Mpc & 
 &
 &
mag &
mag arcsec$^{-2}$ &
 \\

\tableline
\\

NGC 337         & 00:59:50.0 $\;$ -07:34:41 & SB(s)d      &   52                                                 & 130                             & $2.9  \times 1.8  $ & $19.5\pm1.5 $ & TF    & 95  &  $-22.35  \pm 0.18 $  & 21.54  &   1  \\
NGC 628         & 01:36:41.7 $\;$ 15:47:00   & SA(s)c      &   8.7\tablenotemark{f}     & 21\tablenotemark{f}  & $10.5 \times 9.5  $ & $8.6 \pm0.3 $ & PNLF  & 42  &  $-22.83  \pm 0.09 $  & 22.56  &   3  \\
NGC 1042       & 02:40:24.0 $\;$ -08:26:02 & SAB(rs)cd &   40                                                 & 6\tablenotemark{c}    & $4.7  \times 3.6  $ & $4.2 \pm0.7 $ & TF    & 20  &  $-19.27  \pm 0.36 $  & 23.27  &   2  \\
NGC 1068       & 02:42:40.2 $\;$ -00:00:48 & SA(rs)b    &   32                                                 & 70                               & $7.1  \times 6.0  $ & $10.1\pm1.7 $ & TF    & 49  &  $-24.23  \pm 0.36 $  & 19.54  &   3  \\
NGC 2775       & 09:10:20.1 $\;$ 07:02:17   & SA(r)ab    &   40                                                 & 155                              & $4.3  \times 3.3  $ & $21.5\pm1.5 $ & Flow  & 104 &  $-24.60  \pm 0.15 $  & 20.94  &   3  \\
NGC 2841       & 09:22:01.8 $\;$ 50:58:31   & SA(r)b      &   67                                                 & 147                              & $8.1  \times 3.5  $ & $14.1\pm1.5 $ & Ceph  & 68  &  $-24.69  \pm 0.23 $  & 21.43  &   3  \\
NGC 2903       & 09:32:09.7 $\;$ 21:30:02   & SB(s)d      &   64                                                 & 17                               & $12.6 \times 6.0  $ & $8.6\pm 1.4 $ & TF    & 41  &  $-23.62  \pm 0.35 $  & 21.31  &   3  \\
NGC 3147       & 10:16:53.2 $\;$ 73:24:04   & SA(rs)bc   &   28                                                 & 155                              & $3.9  \times 3.5  $ & $43.1\pm3.0 $ & Flow  & 209 &  $-25.76  \pm 0.15 $  & 21.16  &   2  \\
NGC 3166       & 10:13:45.0 $\;$ 03:25:31   & SAB(rs)0   &   63                                                 & 87                               & $4.8  \times 2.3  $ & $22.0\pm1.5 $ & Flow  & 107 &  $-24.50  \pm 0.15 $  & 20.38  &   2  \\
NGC 3198       & 10:19:54.9 $\;$ 45:33:09   & SB(rs)c     &   70                                                 & 35                               & $8.5  \times 3.3  $ & $13.7\pm0.5 $ & Ceph  & 66  &  $-22.90  \pm 0.09 $  & 22.70  &   1  \\
NGC 3227       & 10:23:31.5 $\;$ 19:51:48   & SAB(s)pec &   49                                                 & 155                              & $5.4  \times 3.6  $ & $20.3\pm1.4 $ & Flow  & 99  &  $-23.90  \pm 0.15 $  & 22.60  &   2  \\
NGC 3351       & 10:43:58.1 $\;$ 11:42:15   & SB(r)b       &   49                                                 & 13                               & $7.4  \times 5.0  $ & $9.6\pm 0.2 $ & TRGB  & 47  &  $-21.97  \pm 0.05 $  & 21.57  &   2  \\
NGC 3521       & 11:05:49.0 $\;$ -00:02:15 & SAB(rs)bc &   64                                                 & 163                              & $11.0 \times 5.1  $ & $11.2\pm1.8 $ & TF    & 54  &  $-24.47  \pm 0.35 $  & 20.69  &   3  \\
NGC 3627       & 11:20:15.0 $\;$ 12:59:29   & SAB(s)b    &   65                                                 & 173                              & $9.1  \times 4.2  $ & $8.3\pm0.3  $ & TRGB  & 40  &  $-23.71  \pm 0.09 $  & 20.84  &   3  \\
NGC 3938       & 11:52:49.8 $\;$ 44:07:26   & SA(s)c      &   25                                                 & 52\tablenotemark{c}   & $5.4  \times 4.9  $ & $15.6\pm1.1 $ & Flow  & 75  &  $-23.15  \pm 0.16 $  & 22.11  &   2  \\
NGC 3949       & 11:53:41.5 $\;$ 47:51:35   & SA(s)bc    &   57                                                 & 120                              & $2.9  \times 1.7  $ & $19.1\pm3.1 $ & TF    & 92  &  $-22.80  \pm 0.35 $  & -      &   1  \\
NGC 4013       & 11:58:31.7 $\;$ 43:56:48   & SAb         &   90                                                 & 66                               & $5.2  \times 1.0  $ & $18.9\pm3.1 $ & TF    & 92  &  $-23.75  \pm 0.35 $  & 22.95  &   2  \\
NGC 4254       & 12:18:49.4 $\;$ 14:25:07   & SA(s)c      &   30                                                 & 60\tablenotemark{c}   & $5.4  \times 4.7  $ & $14.3\pm3.5 $ & TF  & 70  &  $-26.07  \pm 0.15 $  & 21.16  &   3  \\
NCG 4314       & 12:22:32.2 $\;$ 29:53:47   & SB(rs)a     &   28                                                 & 145\tablenotemark{c}  & $4.2  \times 3.7  $ & $9.7\pm3.6 $ & TF  & 87  &  $-23.81  \pm 0.15 $  & 21.11  &   2  \\
NGC 4450       & 12:28:29.4 $\;$ 17:05:05   & SA(s)ab    &   43                                                 & 175                              & $5.2  \times 3.9  $ & $15.3\pm2.5 $ & TF    & 74  &  $-23.87  \pm 0.35 $  & 21.79  &   2  \\
NGC 4569       & 12:36:50.1 $\;$ 13:09:48   & SAB(rs)ab &   65                                                 & 23                               & $9.5  \times 4.4  $ & $9.9\pm0.2  $ & STF   & 48  &  $-23.39  \pm 0.06 $  & 22.10  &   1  \\
NGC 4826       & 12:56:44.3 $\;$ 21:41:05   & SA(rs)ab   &   59                                                 & 115                              & $10.0 \times 5.4  $ & $4.4\pm0.1  $ & TRGB  & 21  &  $-22.87  \pm 0.03 $  & 20.69  &   1  \\
NGC 5055       & 13:15:49.3 $\;$ 42:02:06   & SA(rs)bc   &   57                                                 & 105                              & $12.6 \times 7.2  $ & $9.0\pm0.1  $ & TRGB  & 44  &  $-24.16  \pm 0.03 $  & 21.39  &   1  \\
NGC 5194       & 13:29:53.4 $\;$ 47:11:48   & SA(s)bc    &   20\tablenotemark{b}                     & 163                              & $11.2 \times 6.9  $ & $9.1\pm0.6  $ & Flow  & 44  &  $-24.30  \pm 0.15 $  & 21.40  &   3  \\
NGC 5713       & 14:40:11.6 $\;$ -00:17:26 & SAB(rs)bc &   28                                                 & 10                               & $2.8  \times 2.5  $ & $31.3\pm2.2 $ & Flow  & 152 &  $-24.15  \pm 0.15 $  & 21.36  &   1  \\
NGC 5981       & 15:37:53.4 $\;$ 59:23:34   & Sc	        &   90                                                 & 140                              & $2.8  \times 0.5  $ & $49.7\pm9.2 $ & TF    & 241 &  $-24.19  \pm 0.40 $  & -      &   1  \\
NGC 6503       & 17:49:27.7 $\;$ 70:08:41   & SA(s)cd    &   74                                                 & 123                              & $7.1  \times 2.4  $ & $4.0\pm 0.1 $ & TRGB  & 19  &  $-20.71  \pm 0.04 $  & 21.08  &   2  \\ 
NGC 6946       & 20:34:52.3 $\;$ 60:09:14   & SAB(rs)cd &   32                                                 & 60\tablenotemark{c}   & $11.5 \times 9.8  $ & $6.1\pm0.6  $ & PNLF  & 29  &  $-23.55  \pm 0.21 $  & 22.93  &   2  \\
NGC 7479       & 23:04:57.1 $\;$ 12:19:18   & SB(s)c      &   42                                                 & 25                               & $4.1  \times 3.1  $ & $30.2\pm5.6 $ & TF    & 146 &  $-24.20  \pm 0.40 $  & 22.42  &   2  \\
NGC 7331       & 22:37:04.0 $\;$ 34:24:56   & SA(s)b     &   72                                                 & 171                              & $10.5 \times 3.7  $ & $14.5\pm0.6 $ & Ceph  & 70  &  $-24.78  \pm 0.09 $  & 21.51  &   1  \\
\enddata

\tablenotetext{a}{Coordinates, inclination ($i$), position angle ($\theta$), isophotal diameter ($d_{25}$), and effective $B$-band surface brightness ($\mu_{B}$) taken from RC3 \citep{devaucouleurs91} except when indicated.}
\tablenotetext{b}{For NGC5194 we use kinematic inclination angle derived by \cite{tully74}.}
\tablenotetext{c}{From \cite{paturel00} (NGC1042, NGC3938), \cite{springob09} (NGC4254, NGC6946), and \cite{jarrett03} (NGC4314).}
\tablenotetext{d}{Distance ($D$) methods and references; TRGB: Tip of the red giant branch \citep{jacobs09, tully09} except for NGC3351 taken from \cite{rizzi07}; Ceph: Cepheid variables (\cite{freedman01}, except for NGC2841 taken from \cite{macri01}); TF: HI 21cm Tully-Fisher \citep[][for NGC0337 we used the group Tully-Fisher distance]{tully09} except for NGC4252 taken from \cite{springob09}; STF: Stellar kinematics Tully-Fisher \citep{cortes08}; PNLF: Planetary nebulae luminosity function \citep{herrmann08}; Flow: Derived from redshift, and corrected for peculiar velocities \citep[][taken from NED]{mould00}.}
\tablenotetext{e}{From \cite{jarrett03} except for NGC 1042, NGC 3147, NGC 3949, NGC 5981, NGC 7479, and NGC 7331 taken from \cite{jarrett00}.}
\tablenotetext{f}{From this work.}

\end{deluxetable*}

\clearpage

\begin{deluxetable}{lcc}
\tabletypesize{\scriptsize}
\tablecaption{Bulge Structural Parameters\label{tbl-2}}
\tablewidth{0pt}
\tablehead{
\colhead{\ \ \ \ \ \ Object\ \ \ \ \ \ } & 
\colhead{\ \ \ \ \ \ B/T\ \ \ \ \ \ \ } &
\colhead{\ \ \ \ \ \ $n_{Bulge}$\ \ \ \ \ \ } 
}
\startdata
NGC 337       &  -                     & -     \\
NGC 628       &  0.10\tablenotemark{a} & 1.35  \\
NGC 1042       &  -                     & -     \\
NGC 1068       &  -                     & -     \\
NGC 2775       &  0.61\tablenotemark{b} & 4.85  \\
NGC 2841       &  0.17\tablenotemark{c} & 2.97  \\
NGC 2903       &  0.09\tablenotemark{c} & 0.42  \\
NGC 3147       &  0.25\tablenotemark{a} & 3.66  \\
NGC 3166       &  0.25\tablenotemark{b} & 0.56  \\
NGC 3198       &  0.11\tablenotemark{a} & 5.12  \\
NGC 3227       &  -                     & -     \\
NGC 3351       &  0.17\tablenotemark{c} & 1.51  \\
NGC 3521       &  0.10\tablenotemark{a} & 3.20  \\
NGC 3627       &  0.08\tablenotemark{c} & 2.90  \\
NGC 3938       &  0.07\tablenotemark{b} & 1.18  \\
NGC 3949       &  -                     & -     \\
NGC 4013       &  -                     & -     \\
NGC 4254       &  0.39\tablenotemark{b} & 2.68  \\
NGC 4314       &  -                     & -     \\
NGC 4450       &  0.17\tablenotemark{b} & 2.26  \\
NGC 4569       &  0.06\tablenotemark{c} & 1.90  \\
NGC 4826       &  0.13\tablenotemark{c} & 3.94  \\
NGC 5055       &  0.26\tablenotemark{a} & 1.84  \\
NGC 5194       &  -                     & -     \\
NGC 5713       &  0.33\tablenotemark{b} & 1.84  \\
NGC 5981       &  -                     & -     \\
NGC 6503       &  -                     & -     \\ 
NGC 6946       &  -                     & -     \\
NGC 7479       &  0.09\tablenotemark{b} & 1.09  \\
NGC 7331       &  -                     & -     \\
\enddata

\tablenotetext{a}{K-band Decomposition, Dong \& De Robertis 2006}
\tablenotetext{b}{H-band Decomposition, Weinzirl et al. 2008}
\tablenotetext{c}{V-band Decomposition, Fisher \& Drory 2008}

\end{deluxetable}


\clearpage
\begin{deluxetable}{lccc}
\tabletypesize{\scriptsize}
\tablecaption{Stellar Masses and Star Formation Rates\label{tbl-3}}
\tablewidth{0pt}
\tablehead{
\colhead{Object} & 
\colhead{log($M_*$)} &
\colhead{log($SFR$)} &
\colhead{Ref.} 
}

\startdata
 &
M$_{\odot}$ & 
M$_{\odot}$ yr$^{-1}$ & 
 \\

\tableline
\\

NGC 337       &  10.2  &   0.63    &   K03\tablenotemark{a}  \\
NGC 628       &  10.3  &   0.30    &   L09\tablenotemark{c}  \\
NGC 1042       &  8.9   &   0.15    &   T07\tablenotemark{b}  \\
NGC 1068       &  10.9  &   1.59    &   T07  \\
NGC 2775       &  11.1  &   0.06    &   T07  \\
NGC 2841       &  11.1  &  -0.70    &   K03  \\
NGC 2903       &  10.7  &   0.56    &   L09  \\
NGC 3147       &  11.5  &   -       &   -    \\
NGC 3166       &  11.0  &   -       &   -    \\
NGC 3198       &  10.4  &  -0.07    &   K03  \\
NGC 3227       &  10.8  &   -       &   -    \\
NGC 3351       &  10.0  &   0.20    &   L09  \\
NGC 3521       &  11.0  &   0.38    &   L09  \\
NGC 3627       &  10.7  &   0.69    &   L09  \\
NGC 3938       &  10.5  &   0.08    &   K03  \\
NGC 3949       &  10.3  &   -       &   -    \\
NGC 4013       &  10.7  &   -       &   -    \\
NGC 4254       &  11.6  &   1.04    &   K03  \\
NGC 4314       &  10.7  &  -0.18    &   T07  \\
NGC 4450       &  10.8  &  -0.30    &   K03  \\
NGC 4569       &  10.6  &   0.28    &   K03  \\
NGC 4826       &  10.4  &   0.07    &   L09  \\
NGC 5055       &  10.9  &   0.48    &   L09  \\
NGC 5194       &  10.9  &   0.88    &   L09  \\
NGC 5713       &  10.9  &   0.99    &   T07  \\
NGC 5981       &  10.9  &   -       &   -    \\
NGC 6503       &  9.5   &  -0.50    &   L09  \\ 
NGC 6946       &  10.6  &   0.96    &   L09  \\
NGC 7479       &  10.9  &   1.21    &   T07  \\
NGC 7331       &  11.1  &   0.62    &   K03  \\
\enddata

\tablenotetext{a}{K03: Kennicutt et al. 2003}
\tablenotetext{b}{T07: Thilker et al. 2007}
\tablenotetext{c}{L09: Lee et al. 2009}

\end{deluxetable}

\clearpage

\begin{deluxetable}{cccc}
\tabletypesize{\scriptsize}
\tablecaption{VENGA Observing Runs\label{tbl-4}}
\tablewidth{0pt}
\tablehead{
\colhead{Dates} & 
\colhead{Observed Nights} &
\colhead{Instrumental Setup} &
\colhead{Observed Galaxies} 
}
\startdata
08/04/2008                          &  1   & red  & NGC 5194           \\
11/04/2008 - 11/09/2008  &  6   & red  & NGC 628, NGC 1068  \\
01/28/2009 - 01/31/2009  &  4   & red  & NGC 2903, NGC 3521  \\
02/01/2009 - 02/03/2009  &  3   & red  & NGC 1042, NGC 2775, \\
                         &      &      & NGC 3227, NGC 3949, \\
                         &      &      & NGC 4314           \\
03/30/2009 - 04/02/2009  &  3   & red  & NGC 3351, NGC 4254  \\
04/17/2009 - 04/19/2009  &  4   & red  & NGC 4254, NGC 5194  \\
07/15/2009 - 07/16/2009  &  2   & red  & NGC 5713, NGC 6503  \\
07/21/2009 - 07/23/2009  &  1   & red  & NGC 6503           \\
09/11/2009 - 09/15/2009  &  3   & red  & NGC 337, NGC 1068, \\
                         &      &      & NGC 5981, NGC 6503, \\
                         &      &      & NGC 7479           \\
11/09/2009 - 11/15/2009  &  4   & red  & NGC 628, NGC 1042, \\
                         &      &      & NGC 1068, NGC 2775  \\
12/09/2009 - 12/21/2009  &  12  & red  & NGC 628, NGC 1042, \\
                         &      &      & NGC 2775, NGC 2841, \\
                         &      &      & NGC 3166, NGC 3227, \\
                         &      &      & NGC 3521, NGC 3627  \\
01/11/2010 - 01/16/2010  &  4   & red  & NGC 1042, NGC 2841, \\
                         &      &      & NGC 3147, NGC 3627, \\
                         &      &      & NGC 4013           \\
02/14/2010 - 02/18/2010  &  4   & red  & NGC 1068, NGC 2775, \\
                         &      &      & NGC 3147, NGC 3198, \\
                         &      &      & NGC 4013, NGC 4254  \\
05/18/2010 - 05/20/2010  &  3   & red  & NGC 3998, NGC 5055  \\
06/04/2010 - 06/06/2010  &  2   & red  & NGC 3198, NGC 4450, \\
                         &      &      & NGC 6964           \\
07/05/2010 - 07/09/2010  &  3   & red  & NGC 4450, NGC 6946  \\
09/01/2010 - 09/07/2010  &  6   & blue & NGC 1068, NGC 5981, \\
                         &      &      & NGC 6503, NGC 6946, \\
                         &      &      & NGC 7479, NGC 7731  \\
10/01/2010 - 10/06/2010  &  5   & blue & NGC 337, NGC 628, \\
                         &      &      & NGC 1068, NGC 6946, \\
                         &      &      & NGC 7731           \\
11/10/2010 - 11/14/2010  &  2   & blue & NGC 628           \\
12/07/2010 - 12/12/2010  &  5   & blue & NGC 628, NGC 1042, \\
                         &      &      & NGC 2775, NGC 2903  \\
12/27/2010 - 01/02/2011  &  5   & blue & NGC 628, NGC 2775, \\
                         &      &      & NGC 2841, NGC 3147, \\
                         &      &      & NGC 3227           \\
01/27/2011 - 02/02/2011  &  4   & blue & NGC 628, NGC 2775, \\
                         &      &      & NGC 2841, NGC 3147, \\ 
                         &      &      & NGC 3166, NGC 3198, \\
                         &      &      & NGC 3227           \\
02/07/2011 - 02/10/2011  &  3   & blue & NGC 2841, NGC 3147, \\
                         &      &      & NGC 3166, NGC 3198, \\
                         &      &      & NGC 3227, NGC 3521  \\
03/28/2011 - 03/31/2011  &  4   & blue & NGC 2775, NGC 3147, \\
                         &      &      & NGC 3351, NGC 3627, \\
                         &      &      & NGC 5713           \\
04/08/2011 - 04/10/2011  &  1   & blue & NGC 3351, NGC 3949  \\
05/04/2011 - 05/09/2011  &  5   & blue & NGC 3227, NGC 3949,  \\
                         &      &      &  NGC 3938, NGC 5194, \\
                         &      &      &  NGC 5055, NGC 3521, \\
                         &      &      &  NGC 3521, NGC 4826 \\
06/24/2011 - 07/04/2011  &  5   & blue & NGC 4254, NGC 4826,  \\
                         &      &      &  NGC 4013 \\
03/16/2012 - 03/22/2012  &  4   & blue & NGC 3521, NGC 4013,  \\
                         &      &      & NGC 4314, NGC 4450 \\
                         &      &      & NGC 4569 \\

\enddata
\end{deluxetable}

\clearpage

\clearpage

\begin{deluxetable}{cccccccc}
\tabletypesize{\scriptsize}
\tablecaption{Summary of VENGA Observations of NGC 628\label{tbl-5}}
\tablewidth{0pt}
\tablehead{
\colhead{Pointing} & 
\colhead{Equatorial Coord.} & 
\colhead{Setup} &
\colhead{Dither} &
\colhead{Exposure Time} &
\colhead{N} &
\colhead{$\langle$Seeing$\rangle$} &
\colhead{$\langle$Transparency$\rangle$} 
}
\startdata
 &
\ $\alpha$\ \ \ \ \ \ \ \ \ \ \ \ \ $\delta$ &
 &
 & 
hours &
  &
$''$ & 
\\
\tableline
    & & & & & & &\\
    &                                          & red  & D1  & 4.00  & 12 & 2.06 & 0.87 \\
P1  & 01:36:42.45 15:47:04.6  & red & D2  & 3.33  & 10 & 2.23 & 0.87 \\
      &                                        & red & D3  & 3.33  & 10 & 2.20 & 0.89 \\
      &                                        & blue & D1  & 3.33  & 8 & 2.19 & 0.71 \\
      &                                        & blue & D2  & 2.92  & 7 & 2.06 & 0.64 \\
      &                                        & blue & D3  & 3.33  & 8 & 1.55 & 0.73 \\
    & & & & & & &\\
\tableline
    & & & & & & &\\
    &                                         & red & D1  & 3.00  & 6 & 1.92 & 0.65 \\
P2  & 01:36:49.45 15:47:04.2 & red & D2  & 3.50  & 7 & 2.00 & 0.67 \\
    &                                         & red & D3  & 3.50  & 7 & 1.87 & 0.68 \\
     &                                        & blue & D1  & 5.00  & 12 & 1.72 & 0.68 \\
     &                                        & blue & D2  & 5.00  & 12 & 1.53 & 0.69 \\
     &                                        & blue & D3  & 5.00  & 12 & 2.23 & 0.65 \\
    & & & & & & &\\
\tableline
    & & & & & & &\\
    &                                         & red & D1  & 8.50  & 17 & 2.62 & 0.63 \\
P3  & 01:36:35.51 15:47:05.0 & red & D2  & 7.50  & 15 & 2.72 & 0.61 \\
    &                                         & red & D3  & 7.50  & 15 & 2.57 & 0.68 \\
      &                                        & blue & D1  & 5.42  & 13 & 2.27 & 0.49 \\
      &                                        & blue & D2  & 5.42  & 13 & 2.82 & 0.58 \\
      &                                        & blue & D3  & 4.58  & 11 & 2.18 & 0.58 \\
    & & & & & & &\\
\enddata
\end{deluxetable}

\clearpage

\begin{deluxetable}{ccccc}
\tabletypesize{\scriptsize}
\tablecaption{Fitted Emission Lines in NGC 628\label{tbl-6}}
\tablewidth{0pt}
\tablehead{
\colhead{Transition} & 
\colhead{Wavelength} & 
\colhead{Median S/N} &
\colhead{Fraction - $5\sigma$} &
\colhead{Fraction - $3\sigma$}
}
\startdata
 &
\AA\ &
      &
    & 
\\
\tableline
 &  &  &     & \\

[OII]\tablenotemark{a}               & 3726.03 & 16.4  & 0.99     & 1.0\\

[OII]\tablenotemark{a}                  & 3728.73 & -  & -     & -\\

[NeIII]               & 3868.69 & 1.3  & $<$0.01     & $<$0.01\\

[NeIII]               & 3967.40 & 0.7  & $<$0.01     & $<$0.01\\

H8                    & 3889.06 & 2.0 & 0.13     & 0.27\\

H$\epsilon$     & 3970.08 & 2.7  & 0.18     & 0.42\\

H$\delta$        & 4101.73 & 3.0  & 0.26     & 0.44\\

H$\gamma$    & 4340.47 & 8.0  & 0.74     & 0.89\\

[OIII]                & 4363.15 & 0.9  & $<$0.01     & $<$0.01\\

HeII                 & 4685.74 & 0.4  & $<$0.01     & $<$0.01\\

H$\beta$        & 4861.32 & 23.5 & 0.97  & 0.99\\

[OIII]                & 4958.83 & 3.4  & 0.24   & 0.58\\

[OIII]               & 5006.77 & 9.5  & 0.89  & 0.98\\

[NI]\tablenotemark{a}                  & 5197.90 & 1.9  & 0.01     & 0.15\\
           
[NI]\tablenotemark{a}                  & 5200.39 & -  & -    & - \\

[NII]                & 6547.96 & 11.3  & 0.92  & 0.98\\

H$\alpha$      & 6562.80 & 47.4 & 1.00  & 1.00\\

[NII]                & 6583.34 & 26.2 & 0.99  & 1.00\\

[SII]                 & 6716.31 & 16.5 & 0.98  & 1.00\\

[SII]                & 6730.68 & 10.4  & 0.88  & 0.97\\

\enddata
\tablenotetext{a}{Since we cannot resolve the [OII]$\lambda$3727 and
  [NI]$\lambda$5200 doublets, we report the median S/N and fraction of
  the observed area in which the lines are significantly detected for
  the sum of the two doublet components.}

\end{deluxetable}

\end{document}